\documentclass[pdflatex,sn-mathphys-num]{sn-jnl}

\usepackage{graphicx}%
\usepackage{multirow}%
\usepackage{amsmath,amssymb,amsfonts}%
\usepackage{amsthm}%
\usepackage{mathrsfs}%
\usepackage[title]{appendix}%
\usepackage{textcomp}%
\usepackage{manyfoot}%
\usepackage{booktabs}%
\usepackage{algorithm}%
\usepackage{algorithmicx}%
\usepackage{algpseudocode}%
\usepackage{listings}%
\usepackage{amssymb}
\usepackage{tabularx}

\usepackage[table]{xcolor}

\theoremstyle{thmstyleone}%
\theoremstyle{thmstyletwo}%

\theoremstyle{thmstylethree}%

\begin{document}

\title[Distributional ICU Mortality Modeling in Diabetic heart failure Patients]{Predicting Short-Term Mortality in Elderly ICU Patients with Diabetes and Heart Failure: A Distributional Inference Framework}


\author[1]{\fnm{Junyi} \sur{Fan}}
\author[1]{\fnm{Shuheng} \sur{Chen}}
\author[1]{\fnm{Li} \sur{Sun}}
\author[1]{\fnm{Yong} \sur{Si}}
\author[2]{\fnm{Elham} \sur{Pishgar}}
\author[3]{\fnm{Kamiar} \sur{Alaei}}
\author[4]{\fnm{Greg} \sur{Placencia}}
\author*[1]{\fnm{Maryam} \sur{Pishgar}}\email{pishgar@usc.edu}

\affil*[1]{\orgdiv{Department of Industrial and Systems Engineering}, \orgname{University of Southern California}, \orgaddress{\city{Los Angeles}, \state{CA}, \country{United States}}}

\affil[2]{\orgdiv{Colorectal Research Center}, \orgname{Iran University of Medical Sciences}, \orgaddress{\city{Tehran}, \country{Iran}}}

\affil[3]{\orgdiv{Department of Health Science}, \orgname{California State University}, \orgaddress{\city{Long Beach}, \state{CA}, \country{United States}}}

\affil[4]{\orgdiv{Department of Industrial and Manufacturing Engineering}, \orgname{California State Polytechnic University}, \orgaddress{\city{Pomona}, \state{CA}, \country{United States}}}


\abstract{
Elderly ICU patients with coexisting diabetes mellitus and heart failure experience markedly elevated short-term mortality, yet few predictive models are tailored to this high-risk group. Diabetes mellitus affects nearly 30\% of U.S. adults over 65, and significantly increases the risk of heart failure. When combined, these conditions worsen frailty, renal dysfunction, and hospitalization risk, leading to one-year mortality rates of up to 40\%. Despite their clinical burden and complexity, no established models address individualized mortality prediction in elderly ICU patients with both diabetes mellitus and heart failure.

We developed and validated a probabilistic mortality prediction framework using the MIMIC-IV database, targeting 65-90-year-old patients with both diabetes mellitus and heart failure. Using a two-stage feature selection pipeline and a cohort of 1,478 patients, we identified 19 clinically significant variables that reflect physiology, comorbidities, and intensity of treatment. Among six ML models benchmarked, CatBoost achieved the highest test AUROC (0.863), balancing performance and interpretability. 

To enhance clinical relevance, we employed the DREAM algorithm to generate posterior mortality risk distributions rather than point estimates, enabling assessment of both risk magnitude and uncertainty. This distribution-aware approach facilitates individualized triage in complex ICU settings. 

Interpretability was further supported via ablation and ALE analysis, highlighting key predictors such as APS III, oxygen flow, GCS eye, and Braden Mobility. Our model enables transparent, personalized, and uncertainty-informed decision support for high-risk ICU populations.
}

\keywords{ICU mortality prediction, Elderly patients, Machine learning, diabetes mellitus and heart failure comorbidity, DREAM posterior inference, Probabilistic risk stratification}



\maketitle

\section{Introduction}\label{sec1}

Diabetes mellitus is a chronic metabolic disorder characterized by persistent hyperglycemia due to defects in insulin secretion, insulin action, or both, and is especially prevalent among elderly populations\cite{foster1915diabetes,alam2014general,kaul2013introduction,blair2016diabetes}. According to the U.S. Centers for Disease Control and Prevention, 29.2\% of adults aged 65 years or older are living with diabetes, compared to just 14.7\% among the overall U.S. adult population\cite{cdc2023}. Globally, the International Diabetes Federation (IDF) estimates that by 2024, over 135 million people aged 65–99 years are affected by diabetes, a figure projected to nearly double to 276 million by 2045 as the world’s population ages\cite{idf2025}.

Elderly adults with diabetes frequently experience a wide range of complications, among which heart failure is recognized as both highly prevalent and clinically challenging\cite{papatheodorou2018complications,gregg2016changing}. Epidemiological studies have demonstrated that the incidence of heart failure in this population is approximately two to four times that of their non-diabetic peers, underscoring the significant clinical burden posed by this comorbidity\cite{kannel1979diabetes, bertoni2004heart, verny2007congestive}. Notably, the Framingham Heart Study demonstrated that diabetes conferred a 2.4-fold higher risk of heart failure in men and a fivefold risk in women, independent of other cardiovascular risk factors\cite{kannel1979diabetes}. In the elderly population, the coexistence of heart failure and diabetes is particularly detrimental, with large cohort studies indicating a 50–100\% higher risk of all-cause and cardiovascular mortality compared to elderly patients with heart failure alone\cite{macdonald2008diabetes, shah2017heart, birkeland2020heart}. For instance, Shah et al. reported one-year mortality rates approaching 40\% in older adults hospitalized with both conditions, substantially exceeding rates observed in patients with only heart failure\cite{shah2017heart}.
Managing these comorbidities in elderly patients is especially challenging, as diabetes accelerates the progression of heart failure and exacerbates renal dysfunction, frailty, and hospital admission risk\cite{weng2023effect, strain2021diabetes}. The combined burden also translates into substantially higher healthcare costs, with annual expenses estimated to be two to three times greater for patients with both conditions—driven largely by recurrent hospitalizations, advanced interventions, and long-term care needs\cite{heidenreich2013forecasting, american2024introduction}. Thus, these findings underscore the critical need to develop robust predictive models and identify key prognostic factors to guide risk stratification and support clinical decision-making for elderly ICU patients with coexisting diabetes and heart failure, ultimately aiming to improve outcomes in this high-risk population.

Aguilar et al. (2009) retrospectively analyzed 5,815 diabetic patients with established heart failure from a national Veterans Affairs (VA) cohort, stratifying subjects by quintiles of glycosylated hemoglobin (HbA1C) to assess associations with two-year mortality and heart failure hospitalization. Their findings revealed a U-shaped relationship between HbA1C and mortality: the lowest risk was observed in patients with moderate glycemic control (HbA1C 7.1–7.8\%, adjusted hazard ratio: 0.73, 95\% CI: 0.61–0.88), while both lower and higher HbA1C levels were linked to increased mortality. Although higher HbA1C was associated with increased heart failure hospitalization rates, this was not significant after adjustment. The results identify HbA1C as a key predictor of mortality in this population, with optimal outcomes seen in patients with modest glucose control\cite{aguilar2009relationship}.

Evans et al. (2010) conducted a large population-based cohort study in Tayside, Scotland, to evaluate the impact of metformin therapy on mortality in patients with type 2 diabetes mellitus and chronic heart failure (CHF). Using linked prescribing and clinical databases, 422 patients with diabetes mellitus and incident CHF receiving oral hypoglycemic agents were identified and grouped by treatment: metformin monotherapy, sulfonylurea monotherapy, and combination therapy. Cox regression analysis, adjusted for comorbidities and concurrent therapies, revealed that metformin use—either alone or in combination—was associated with significantly lower all-cause mortality at one year (hazard ratio: 0.59, 95\% CI: 0.36 to 0.96) and over long-term follow-up (hazard ratio: 0.67, 95\% CI: 0.51 to 0.88) compared to sulfonylurea monotherapy. These findings identify metformin therapy as a potential predictor of improved survival in patients with diabetes mellitus and CHF\cite{evans2010effect}.

Zhou et al. (2023) investigated the prognostic value of stress hyperglycemia, as measured by the stress hyperglycemia ratio (SHR), in diabetic patients with acute decompensated heart failure (ADHF). In this cohort study of 780 diabetic ADHF patients from a larger group of 1,904, SHR was calculated as admission blood glucose divided by estimated average glucose from glycosylated hemoglobin (HbA1C). Patients were stratified by SHR quintiles and followed for a median of 3.24 years. Using Cox proportional hazards and restricted cubic spline analyses, the study identified a U-shaped relationship between SHR and all-cause mortality, cardiovascular death, and heart failure rehospitalization at three years. Both the lowest and highest SHR quintiles were associated with significantly increased risks of adverse outcomes compared to the second quintile, which had the lowest mortality. The U-shaped association remained significant across heart failure with preserved ejection fraction (HFpEF) and heart failure with reduced ejection fraction (HFrEF) subgroups. These findings highlight the SHR as an important predictor of long-term prognosis in diabetic patients hospitalized with ADHF\cite{zhou2023impact}.

While previous studies have investigated mortality prediction in patients with diabetes mellitus and heart failure using traditional statistical approaches such as Cox regression or logistic regression, the majority of these investigations have focused on general adult or hospitalized populations, with few specifically addressing elderly patients admitted to the intensive care unit (ICU)\cite{aguilar2009relationship, evans2010effect, zhou2023impact}. Furthermore, although conventional models have identified several important clinical predictors, their ability to account for complex, nonlinear relationships among high-dimensional variables is inherently limited. Given that elderly ICU patients with coexisting diabetes mellitus and heart failure represent a particularly high-risk and clinically challenging subgroup, there is a pressing need for more accurate and individualized risk prediction tools to guide management and optimize outcomes. 

The advent of large-scale electronic health records and advances in computational methods, such as machine learning (ML), now offer a unique opportunity to address this gap by enabling the integration of large-scale clinical data and the identification of novel prognostic patterns. Recent years have witnessed a rapid expansion in the use of ML models for clinical risk prediction and outcome modeling. For example, Si et al.(2025) conducted a study utilized the MIMIC-IV database to develop and validate a CatBoost-based model for predicting in-hospital mortality among intensive care unit (ICU) patients. After rigorous feature selection and benchmarking against other machine learning algorithms, the CatBoost model achieved an AUC of 0.904 in the training set and 0.868 in the test set, demonstrating excellent discriminative ability and potential for integration into clinical risk assessment workflows\cite{si2025retrospective}. Chen et al. (2025) developed an model based on XGBoost to predict the risk of readmission from intensive care unit (ICU) in patients with acute pancreatitis (AP) using data from the MIMIC-III database. After rigorous feature selection and model optimization, their model achieved an Area Under the Receiver Operating Characteristic curve(AUROC) of 0.862 (95\% Confidence Interval (CI): 0.800–0.920) and an accuracy of 0.889 (95\% CI: 0.858–0.923) on the test set. Platelet count, age, and peripheral oxygen saturation (SpO2) were identified as the most important predictors, demonstrating the model’s potential to guide targeted post-discharge interventions for patients with AP.\cite{chen2025predicting}. Gao et al. (2024) developed a random forest-based model to predict outcomes in adult sepsis patients using the MIMIC-IV database. The model, optimized through advanced hyperparameter tuning and rigorous feature selection, achieved an AUROC of 0.94 (±0.01), outperforming other machine learning algorithms. Sequential Organ Failure Assessment (SOFA) score and average urine output emerged as key predictors, and SHAP analysis was used to enhance model interpretability, offering valuable clinical insights for sepsis prognosis\cite{gao2024prediction}. Zhao et al. (2021) developed and prospectively validated a CatBoost-based model to predict extubation failure (EF) in intensive care unit (ICU) patients using data from the MIMIC-IV database. Nineteen key clinical and laboratory characteristics were selected using elimination of recursive characteristics, with duration of mechanical ventilation and pressure support ventilation level identified as the most important predictors by SHapley Additive Explanations (SHAP) analysis. The model achieved an AUROC of 0.835 in internal validation and 0.803 in prospective external validation, demonstrating superior predictive performance compared to other machine learning algorithms\cite{zhao2021development}.
Among these, CatBoost has emerged as a promising algorithm due to several methodological advantages: (1) it efficiently handles categorical variables by employing native encoding strategies, thereby minimizing the need for extensive data preprocessing; (2) it incorporates advanced regularization and ordered boosting techniques, which substantially reduce the risk of overfitting; and (3) it consistently delivers robust predictive performance, even in datasets characterized by high dimensionality, missing values, or relatively small sample sizes. These features make CatBoost especially suitable for the analysis of complex clinical datasets.

This study introduces six key innovations in the development and application of machine learning models to predict mortality at 28 days in elderly ICU patients with co-existing diabetes mellitus and heart failure, addressing important gaps in previous research and advancing clinical and methodological practice.

\begin{itemize}
\item Urgent Clinical Need: To our knowledge, this is the first study to develop and validate a mortality prediction model specifically tailored to elderly ICU patients with both diabetes mellitus and heart failure, using a large, real-world cohort from an open-access database. Existing models rarely address this high-risk group as a distinct population, despite their markedly elevated mortality and complex care requirements
.

\item Posterior-Based Risk Stratification: This study leverages the Density Ratio Estimation for Accurate Modeling(DREAM) algorithm to estimate full mortality risk distributions for elderly ICU patients with diabetes and heart failure. By conditioning on priors from non-survivors, DREAM outputs individualized probability curves rather than fixed scores. These distributions allow clinicians to assess both risk level and uncertainty, supporting more informed, threshold-free decisions. This enhances interpretability and enables personalized triage in complex ICU settings
.

\item Rigorous Feature Selection: We designed a two-stage feature selection pipeline, first applying the Analysis of Variance(ANOVA) F-test to filter over 600 candidate variables, followed by Random Forest Gini importance ranking to refine a clinically actionable set of 19 features. This approach ensured both statistical robustness and clinical interpretability, incorporating expertise from critical care clinicians and targeting key domains such as laboratory results, bedside assessments, procedures, comorbidities, and demographics
.

\item Robust Handling of Imbalanced Data and Validation: To address the pronounced class imbalance typical of ICU mortality, we implemented Synthetic Minority Over-sampling Technique(SMOTE) based oversampling strictly within training folds of a stratified 5-fold cross-validation framework. All data imputation, scaling, and encoding were also performed exclusively within the training data, minimizing information leakage and enhancing generalizability—a methodological rigor not consistently applied in prior ICU studies
.

\item Systematic Model Benchmarking and Interpretability: We conducted a head-to-head comparison of six state-of-the-art machine learning algorithms, including CatBoost, XGBoost, LightGBM, logistic regression, Naive Bayes, and neural networks. CatBoost achieved the highest AUROC on the test set (0.863, 95\% CI: 0.823–0.905) while preserving model interpretability through ablation and Accumulated Local Effects (ALE) analysis, supporting transparent clinical translation and robust individualized risk estimation
.

\item Identification of Dominant Predictors: Through integrated ablation and ALE analysis, we identified the APS III score, oxygen flow, GCS Eye Opening, Braden Mobility, International Normalized Ratio (INR, based on prothrombin time [PT]) and red cell distribution width - standard deviation (RDW-SD) as the most influential predictors of mortality at 28 days. These characteristics collectively reflect the severity of the disease, organ dysfunction, frailty, and respiratory status, allowing pathophysiologically meaningful and actionable risk stratification for clinical decision support at the bedside.
\end{itemize}

\section{Methodology}
\subsection*{Data Source and study design}

This study utilized the MIMIC-IV database \cite{johnson2020mimic}, a publicly available critical care dataset developed by the Massachusetts Institute of Technology in collaboration with the Beth Israel Deaconess Medical Center. The dataset comprises de-identified electronic health records from over 60{,}000 ICU admissions between 2008 and 2019, and has been extensively validated in critical care research \cite{pang2022establishment}.

We proposed a structured, reproducible, and clinically oriented machine learning pipeline to predict in-hospital mortality in elderly ICU patients (aged 65--90) with coexisting diabetes mellitus and heart failure. A total of 35{,}794 patients with first ICU admission were identified, from which 6{,}251 had diabetes, and 2{,}435 were further diagnosed with heart failure. After excluding patients outside the target age range or with active malignancy, the final cohort included 1{,}478 patients.

More than 600 candidate characteristics were initially extracted from five clinical domains: chart events, labe events, drug administration, procedure events, comorbidities, and demographics. Based on recommendations from domain experts Dr. Kamiar Alaei and Dr. Greg Placencia, features were retained only if they had less than 20\% missingness and were documented in at least 100 patients. Imputation was performed using the median for continuous variables and the mode for categorical variables. All continuous features were subsequently standardized using z-score normalization.

To incorporate dynamic trends during the first 24 hours of admission to the ICU, we calculated statistical dispersion metrics that included the coefficient of variation and the interquartile range (IQR) for the selected variables. Feature selection was performed in two stages: initial filtering via ANOVA F-statistics, followed by refinement using Random Forest--based importance ranking. This process yielded 19 final features across all five clinical domains.

To mitigate class imbalance in mortality outcomes, SMOTE was applied solely to the training folds of a stratified 5-fold cross-validation pipeline. Six machine learning models---CatBoost, XGBoost, LightGBM, logistic regression, neural networks, and Na\"ive Bayes---were trained and optimized using GridSearchCV. The performance of the model was evaluated using AUROC, sensitivity, specificity, accuracy, and the F1 score.

Interpretability was enhanced using ablation studies to quantify individual feature contributions. To further support personalized risk communication, posterior distributions of mortality risk were generated using DREAM sampling algorithms, offering 95\% credible intervals for individualized predictions. The overall framework is summarized in Algorithm~\ref{alg:mortality_prediction}, supporting transparent, clinically relevant risk stratification for ICU patients with complex chronic conditions.

\begin{algorithm}[H]
\caption{ML Pipeline for In-Hospital Mortality Prediction in ICU Patients with Diabetes and Heart Failure}
\label{alg:mortality_prediction}
\begin{algorithmic}[1]
\Require MIMIC-IV ICU data with confirmed diagnoses of diabetes and heart failure
\Ensure Binary outcome: in-hospital mortality (0 = survived, 1 = died)

\State \textbf{Step 1: Patient Selection}
\State Extract patients with first ICU admission (\(n = 35{,}794\))
\State Filter by diabetes diagnosis codes (\(n = 6{,}251\))
\State Further filter by heart failure diagnosis (\(n = 2{,}435\))
\State Exclude patients outside 65--90 age range or with malignancy
\State Final cohort: \(n = 1{,}478\)

\State \textbf{Step 2: Data Preprocessing}
\State Remove features with $>$20\% missingness or documented in $<$100 patients
\State Impute numeric values (e.g., \textit{Urea Nitrogen}) using median
\State Impute categorical values (e.g., \textit{Braden Mobility}) using mode
\State Compute coefficient of variation and IQR for time-series variables
\State Normalize continuous features via z-score transformation

\State \textbf{Step 3: Feature Selection}
\State Apply ANOVA F-test to select top 30 features
\State Rank via Random Forest Gini importance
\State Retain 19 clinically interpretable features

\State \textbf{Step 4: Data Balancing}
\State Apply SMOTE to oversample minority class (mortality = 1) in training data only

\State \textbf{Step 5: Model Development}
\State Stratify-split dataset: 70\% training, 30\% test
\ForAll{model $\in$ \{CatBoost, XGBoost, LightGBM, LR, NN, NB\}}
    \State Tune hyperparameters via GridSearchCV
    \State Evaluate via AUROC, accuracy, F1, sensitivity, specificity
\EndFor

\State \textbf{Step 6: Statistical and Interpretability Analysis}
\State Perform t-test on training vs. test distributions
\State Conduct ablation study to assess feature contribution

\State \textbf{Step 7: Posterior Inference}
\State Apply DREAM sampling to compute posterior risk distribution
\State Output 95\% credible interval for individual mortality risk
\end{algorithmic}
\end{algorithm}

\subsection*{Patient selection}

This study aimed to develop a clinically interpretable model to predict 28-day in-hospital mortality among elderly ICU patients (age 65–90) with coexisting diabetes mellitus and heart failure. To construct a well-defined and clinically relevant cohort, we applied a series of structured inclusion and exclusion criteria. Each filtering step was designed to enhance population homogeneity, minimize confounding, and align the final cohort with the intended use case of stratification of risk of geriatric mortality in critical care.

The initial sample included 35,794 patients corresponding to their first ICU admission. Restricting to the first ICU stay per patient mitigates duplication bias and ensures that extracted physiological features reflect acute critical illness rather than chronic readmission-related patterns.

From this group, we identified patients with a documented diagnosis of diabetes mellitus (n = 6,251), based on standardized ICD-9 and ICD-10 codes. Diabetes has well-established associations with impaired immune response, multiorgan vulnerability, and poor recovery in ICU settings. Including this subgroup targets a high-risk population commonly encountered in aging cohorts.

Next, we selected patients with comorbid heart failure (n = 2,435), given its known impact on ICU prognosis through mechanisms such as fluid overload, hemodynamic instability, and limited physiologic reserve. The coexistence of diabetes and heart failure is particularly concerning in older adults and represents a clinically important subgroup for targeted risk prediction.

To define a geriatric cohort, we restricted patient age to between 65 and 90 years at ICU admission (n = 1,759). This range focuses on elderly patients while excluding extreme age outliers who may follow atypical clinical trajectories or present ethical considerations for aggressive ICU care. This age filter enhances model relevance to real-world clinical decision-making in aging populations.

Finally, patients with active malignancy or metastatic cancer were excluded (n = 1,478) to remove individuals whose ICU management may be of palliative intention or subject to limitations of non-curative care. These patients represent a fundamentally different clinical context and could introduce bias into mortality modeling.

The resulting cohort comprised 1,478 elderly ICU patients with coexisting diabetes mellitus and heart failure, free of active malignancy, and experiencing their first ICU admission. For each patient, structured clinical data was extracted from the first 24 hours of stay in the ICU, including demographics, comorbidity profiles, laboratory measurements, vital signs, procedures, and treatments. The primary endpoint was hospital mortality at 28 days, defined according to hospital discharge records and time to death.

\begin{figure}[H]
\centering
\includegraphics[width=0.65\linewidth]{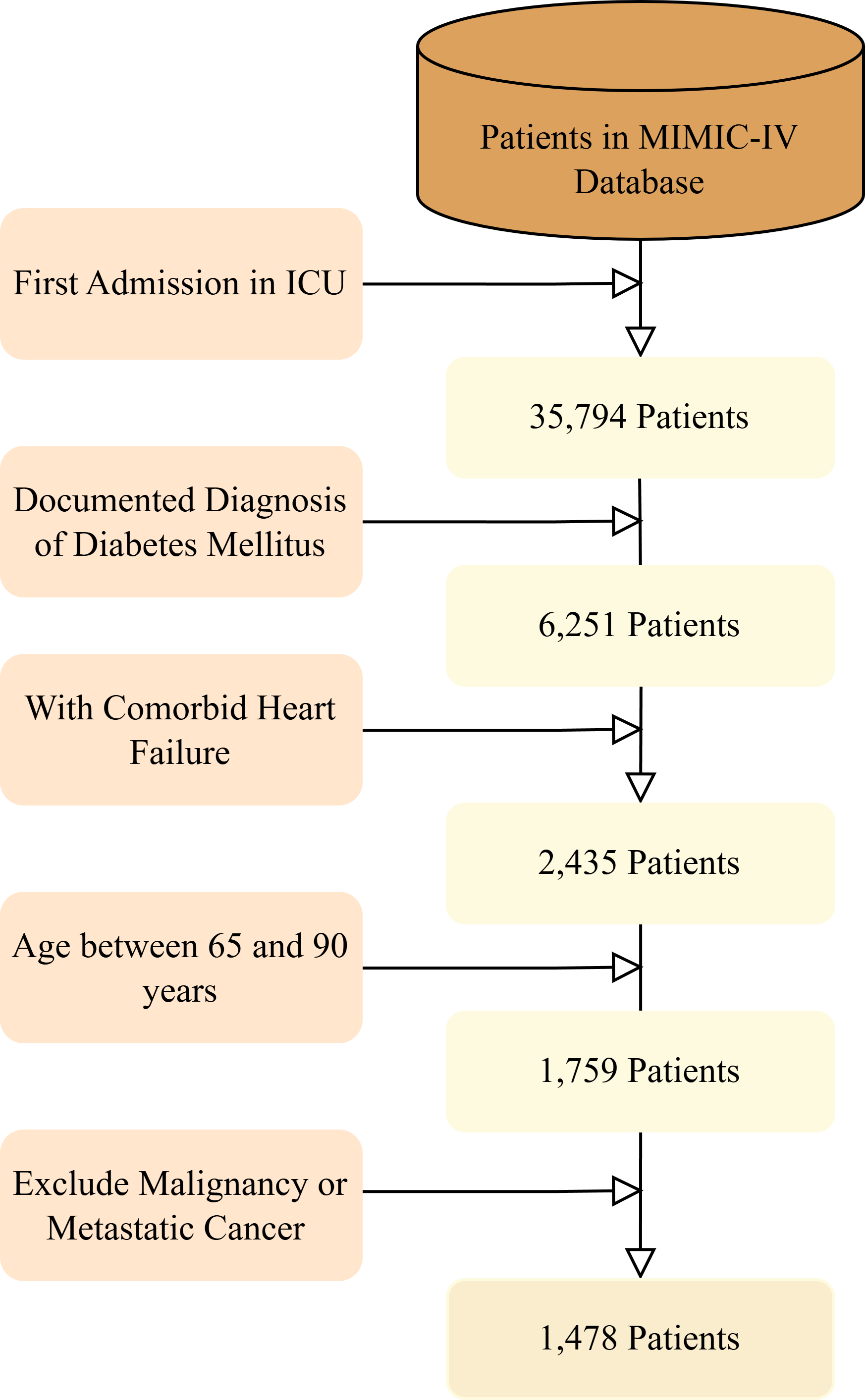} 
\caption{Flow diagram depicting patient inclusion criteria and cohort extraction from the MIMIC-IV database}
\label{fig:patient_selection}
\end{figure}

A visual summary of the cohort selection process is shown in Figure~\ref{fig:patient_selection}.

\subsection*{Data Preprocessing}

To optimize model performance while ensuring clinical interpretability, a structured preprocessing pipeline was applied to MIMIC-IV data. This pipeline emphasized realistic imputation, temporal trend summarization, and model-compatible scaling.

Missing values were imputed using a Random Forest–based technique (MissForest), which iteratively estimates missing entries by fitting decision trees on observed feature subsets. This approach is well-suited for heterogeneous ICU data with complex cross-variable dependencies, and was particularly effective for variables like \textit{RDW-SD}, \textit{Phosphorous}, and \textit{INR (PT)}.

For time-series variables recorded over the first 24 hours of ICU stay, we extracted the minimum, maximum, and slope (trend) for each patient. The slope was computed by fitting a simple linear regression line across hourly values, capturing directional trends in markers such as \textit{O\textsubscript{2} Flow} or \textit{Braden Mobility}:

\begin{equation}
\text{Slope}(x_i) = \frac{d x_i}{dt} \approx \text{coeff}_{1} \text{ in } x_i(t) = \text{coeff}_0 + \text{coeff}_1 \cdot t
\end{equation}

Categorical features were encoded using smoothed target encoding\cite{pargent2022regularized}, mapping each category to its mean mortality rate in the training data. This strategy helped preserve predictive signal while avoiding excessive sparsity from one-hot encoding, especially for features like \textit{Transthoracic Echo} and \textit{Insurance Type}.

All continuous variables were scaled using Min-Max normalization to a 0–1 range:

\begin{equation}
x_i^{*} = \frac{x_i - \min(x_i)}{\max(x_i) - \min(x_i)}
\end{equation}

This ensured that all features were on comparable scales, which benefits distance-based classifiers and facilitates convergence in gradient-boosting algorithms.

Class imbalance in the outcome variable was addressed using SMOTE\cite{si2025machine}, which generates synthetic instances of the minority class by interpolating between existing samples in the feature space. Unlike random oversampling, SMOTE maintains distributional continuity and avoids exact duplication, thus reducing overfitting risk. This technique enhanced the model’s sensitivity to mortality events, particularly in underrepresented high-risk patients.

All resampling and transformation procedures—including imputation, normalization, and encoding—were confined strictly to the training folds within a stratified 5-fold cross-validation framework. Validation and test sets remained untouched to preserve the integrity of performance evaluation and prevent information leakage.

Together, this preprocessing framework enabled a clean, temporally aware, and clinically representative feature matrix suitable for robust supervised learning.

\subsection*{Feature Selection}

We began with a diverse pool of over 600 candidate variables, systematically categorized into five major sources in the MIMIC-IV database. Initial inclusion criteria required variables to meet minimum coverage thresholds—such as documentation for at least 100 patients—and, for medications, a prevalence of 20\% to 50\% across the cohort. This filtering step was designed to exclude extremely sparse, infrequently used, or clinically unreliable variables that may introduce statistical noise or risk model overfitting.

Importantly, the variable inclusion strategy was refined through expert consultation with experienced clinicians in critical care and translational medicine. Their input ensured that selected features not only demonstrated statistical relevance but also possessed diagnostic plausibility, stable documentation patterns, and real-world interpretability in intensive care settings.

The categorization of variables into domains—chartevents, labevents, drug administration, comorbidities, and administrative data were clinically informed. For example, charted bedside scores such as Braden Mobility \cite{JENTZER20191994} capture nursing assessments related to immobility and pressure ulcer risk, which are associated with poor outcomes in heart failure patients. Laboratory markers such as \textit{Lactate} and \textit{Urea Nitrogen} reflect metabolic and renal stress\cite{AZUSHIMA20231135}, both of which are known to worsen prognosis in diabetics with acute decompensation. Comorbidity flags (e.g., \textit{Acute respiratory failure with hypoxia}) directly capture underlying organ dysfunction\cite{vincent2006ali_mods}, and demographic or administrative features (e.g., \textit{Insurance status}, \textit{ED duration}) may influence triage, treatment timelines, and equity of care\cite{shu2023development}.

This clinically grounded categorization improved interpretability, reduced redundancy, and facilitated structured feature engineering in downstream modeling.

\begin{itemize}
    \item \textbf{Chartevents} (n $\approx$ 200): Captured dynamic bedside assessments such as \textit{Braden Mobility}, \textit{Anion gap}, and \textit{Total Bilirubin}.
    
    \item \textbf{Labevents} (n $\approx$ 200): Included laboratory metrics like \textit{Lactate}, \textit{Urea Nitrogen}, and \textit{RDW-SD}.
    
    \item \textbf{Procedure Events (n $\geq$ 100):} Invasive interventions recorded for at least 100 patients, such as \textit{Multi Lumen catheterization}, \textit{Invasive Ventilation}, and \textit{Transthoracic Echocardiography}.
    
    \begin{sloppypar}
    \item \textbf{Drug Administration (20--50\% coverage):} Medications including \textit{Ramelteon} and \textit{Torsemide} were selected based on intermediate prevalence.
    \end{sloppypar}
    
    \item \textbf{Comorbidities} (n = 100): Diagnoses derived from ICD-10, such as \textit{Acute respiratory failure with hypoxia}, were encoded using established clinical groupings.
    
    \item \textbf{Demographics and Administrative} data: Variables such as \textit{Age}, \textit{Insurance Type}, and \textit{ED Duration} were considered to capture variations in access to care and triage outcomes.
\end{itemize}

Variables with excessive missingness ($>$20\%) were removed prior to selection. This resulted in the exclusion of high-cardinality but sparsely documented variables in chartevents and labevents, such as advanced metabolic markers, infrequently used sedation scores, and experimental medication classes.

We applied the ANOVA\cite{dou2023association} F-test using the SelectKBest method to retain the top 30 features showing the highest statistical association with the binary mortality outcome:

\begin{equation}
F(x_i) = \frac{\text{Between-group variance of } x_i}{\text{Within-group variance of } x_i}
\label{eq:anova}
\end{equation}

This step served to eliminate uninformative or redundant variables from the initial pool, reducing dimensionality while preserving clinical breadth. Features retained at this stage primarily included routinely collected laboratory measurements, vital signs recorded within the first 24 hours of ICU admission, and key components from established scoring systems such as SOFA. Demographic variables and early respiratory parameters also demonstrated moderate statistical association with outcome.

A secondary selection was performed using feature importances derived from a trained Random Forest classifier. Importance scores were computed using the Gini impurity reduction criterion:

\begin{equation}
I(x_i) = \sum_{t \in T} \frac{p(t) \cdot \Delta i(t)}{f(t)}
\label{eq:gini}
\end{equation}

where \( I(x_i) \) is the importance of feature \( x_i \), \( T \) is the set of all tree nodes, \( p(t) \) is the proportion of the sample reaching node \( t \), \( \Delta i(t) \) is the decrease in Gini impurity at node \( t \), and \( f(t) \) is the frequency of feature \( x_i \) being used for a split.

In this step, variables with low importance or overlapping information such as co-occurring lab variants, duplicate chart-derived scores, or highly correlated medication administrations—were removed to enhance interpretability and model robustness\cite{zhang2022prediction}.

The final set of 19 characteristics was distributed across six clinically and operationally significant domains, reflecting a balance between physiological signal, intensity of treatment, status of comorbidity, and demographic context. This multidimensional structure was instrumental in capturing the complex determinants of 28-day mortality among elderly ICU patients with comorbid heart failure and diabetes.

Specifically, variables from the \textit{chartevents} domain captured bedside functional assessments (e.g., \textit{Braden Mobility}, \textit{Braden Friction/Shear}) and biochemical markers of systemic dysregulation (e.g., \textit{Anion gap}, \textit{Total Bilirubin}, \textit{GCS - Eye Opening}). Procedure-related variables such as \textit{Invasive Ventilation} and \textit{Transthoracic Echocardiography} reflected acute therapeutic interventions and the complexity of the disease. Laboratory measurements—including \textit{pO\textsubscript{2}}, \textit{INR (PT)}, and \textit{APS III}—served as surrogate indicators of organ failure, coagulopathy, and severity scoring, respectively.

Pharmacological exposures like \textit{Ramelteon} and \textit{Torsemide} represented proxies for sedation management and fluid status control. Meanwhile, critical comorbidities such as \textit{Severe sepsis with septic shock} provided anchoring insight into baseline vulnerability. Demographic attributes such as \textit{Age} were retained for their established prognostic value in geriatric ICU populations.

The final set of characteristics, shown in Table~\ref{tab:final_features} was finalized with the guidance of senior clinical experts, whose perspectives helped prioritize variables that aligned with bedside practices and critical care decision-making. Their contributions were particularly instrumental in validating the inclusion of frailty-related assessments (e.g. Braden subscales), ventilatory interventions, and high-risk comorbidity indicators, ensuring that the resulting model remains both interpretable and actionable in practical ICU workflows.

\begin{table}[H]
\centering
\caption{\textbf{Final 19 Features Used for Mortality Prediction in ICU Patients with Heart Failure and Diabetes}}
\label{tab:final_features}
\small
\renewcommand{\arraystretch}{1.2}
\begin{tabular}{p{0.25\textwidth}|p{0.70\textwidth}}
\hline
\rowcolor[HTML]{E3D5CA}
\textbf{Category} & \textbf{Selected Features} \\
\hline
Chartevents & Anion gap, O\textsubscript{2} Flow, Braden Nutrition, Total Bilirubin, GCS - Eye Opening, Phosphorous, Braden Friction/Shear, Braden Mobility \\
\hline
Procedure Events & Multi Lumen, Invasive Ventilation, Transthoracic Echo \\
\hline
Lab Events & pO\textsubscript{2}, INR(PT), RDW-SD, APS III \\
\hline
Drug Administration & Ramelteon, Torsemide \\
\hline
Comorbidities & Severe sepsis with septic shock \\
\hline
Demographics & Age \\
\hline
\end{tabular}
\end{table}

\subsection*{Basic Model Development}

Following completion of feature engineering, imputation, and variable selection, we constructed a supervised learning framework to predict in-hospital mortality in ICU patients with concurrent heart failure and diabetes. To ensure class balance between subsets, the data was divided into training (70\%) and testing (30\%) cohorts using stratified random sampling, maintaining the original outcome distribution throughout model development.

To ensure robustness, generalizability, and fair comparison, we implemented six machine learning algorithms spanning multiple model families: ensemble-based learners (CatBoost, XGBoost, LightGBM), linear classifiers (Logistic Regression), probabilistic approaches (Naive Bayes), and neural networks (shallow feedforward). Each model was hyperparameter tuned using a 5-fold stratified cross-validation combined with grid search optimization. The independent test set was held entirely from model training and validation to ensure unbiased generalization estimates.

In CatBoost, we employed its ordered boosting scheme and symmetric tree construction, which make it particularly suitable for datasets with moderate size and mixed variable types, as often seen in clinical settings\cite{yuan2024xgboost}. The key hyperparameters included the learning rate, the depth of the tree, the number of booster iterations, and the regularization strength of L2. CatBoost’s ability to handle categorical characteristics natively and reduce overfitting through efficient regularization made it a strong candidate to capture complex non-linear patterns in heterogeneous patient profiles in the ICU.

LightGBM and XGBoost, two additional gradient boosting implementations, were optimized to balance speed, complexity, and performance. For LightGBM, histogram-based decision trees and leaf-wise growth policies accelerated training while maintaining accuracy. For both models, we tuned the learning rate, tree depth, number of estimators, minimum child weight, subsample ratio, column sampling rate (\texttt{colsample\_bytree}), and regularization parameters (\( \text{reg\_alpha}, \text{reg\_lambda} \)). These boosting models are well-suited for high-dimensional datasets with sparse or missing data, making them ideal for clinical mortality prediction where not all lab or chart variables are present for every patient.

The Logistic Regression model served as a strong linear baseline\cite{xu2022predicting}. Both L1 (Lasso) and L2 (Ridge) regularization variants were tested to manage multicollinearity and enforce sparsity. The inverse regularization strength \( C \) was tuned to control the complexity of the model. Despite its simplicity, Logistic Regression offers high interpretability, which is especially important in healthcare applications where transparent decision-making is valued by clinicians.

Naïve Bayes\cite{ren2024prediction} was implemented using a Gaussian assumption for continuous variables. Although this method assumes feature independence, its closed-form solutions and minimal training requirements make it robust under small-sample or high-missingness conditions. It served as a fast, lightweight benchmark to contrast with more complex models and remained stable even under class imbalance when class priors were properly calibrated.

Neural networks were employed to explore nonlinear interactions among clinical features that may not be easily captured by tree-based or distance-based algorithms. A shallow architecture with a single hidden layer was adopted, consistent with previous recommendations~\cite{ding2021artificial}. Key hyperparameters, including the number of hidden units, learning rate, batch size, activation functions (ReLU in hidden layers, sigmoid at the output), and optimizer (Adam)---were tuned using grid search to ensure stable convergence and minimized loss. Although neural networks are inherently less interpretable, their ability to approximate complex functional mappings offers added value in high-dimensional clinical prediction tasks.

Taken together, this ensemble of models spans a spectrum of complexity, interpretability, and computational cost. Their complementary strengths enabled a comprehensive evaluation of linear and non-linear predictive patterns within ICU mortality risk modeling.

Each model was trained with the objective of maximizing classification performance while maintaining generalizability and minimizing overfitting. Loss functions were chosen based on the nature of each algorithm, such as binary cross-entropy for probabilistic models and log loss for gradient boosting frameworks. To further prevent overfitting, regularization techniques were applied where applicable, including L1/L2 penalties, dropout (in neural networks) and tree-level constraints in boosting models.

Model performance was primarily evaluated using the AUROC, which quantifies the model's ability to distinguish between survivors and non-survivors across all classification thresholds. To assess robustness and statistical confidence, 2,000 bootstrap replicas were performed to estimate 95\% CI for AUROC. 

For ensemble models such as CatBoost, XGBoost, and LightGBM, additional variance reduction strategies were used, such as feature subsampling, row sampling, and bagging, to enhance model stability and reduce the risk of overfitting to noise in the training data.

This comprehensive modeling framework, which integrates diverse algorithmic paradigms, rigorous hyperparameter tuning, and statistically grounded evaluation procedures, establishes a solid foundation for developing clinically deployable decision-support tools to stratify the risk of inhospital mortality among diabetic heart failure patients in the ICU.

\subsection*{Distributional Model Development}

To enhance the interpretability of the model beyond single-point probability estimates, we incorporated a Bayesian posterior predictive framework that estimates the full distribution of mortality risk, enabling a more nuanced understanding of prediction uncertainty and its clinical implications. This shift is especially relevant in the prognosis of the ICU, where binary predictions can overlook the nuanced spectrum of patient trajectories. To achieve this, we used the DREAM algorithm~\cite{vrugt2009accelerating}, a distribution-aware framework designed to simulate individualized risk using posterior sampling.

To simulate individualized mortality risk grounded in realistic ICU scenarios, DREAM estimates a full posterior predictive distribution by marginalizing over feature uncertainty. For a patient-level prediction, rather than conditioning on a fixed input vector \( x \), the model integrates over a distribution of plausible high-risk profiles drawn from non-survivor priors:

\begin{equation}
P(\hat{y} \mid \mathcal{D}) = \int P(\hat{y} \mid x, \mathcal{D}) \cdot P(x \mid \text{non-survivor}) \, dx,
\label{eq:dream}
\end{equation}

where \( \mathcal{D} \) is the training dataset used to construct the predictive model \( f(x) = \mathbb{E}[\hat{y} \mid x] \), and \( P(x \mid \text{non-survivor}) \) reflects the empirical distribution of features observed in patients who died within 28 days. The integral is approximated via Monte Carlo sampling, yielding:

\begin{equation}
\hat{P}(\hat{y}) \approx \frac{1}{N} \sum_{i=1}^{N} f(x_i), \quad x_i \sim P(x \mid \text{non-survivor}).
\label{eq:dream_mc}
\end{equation}

This approach enables probabilistic reasoning over the entire mortality distribution for representative patients, accounting for physiological variability and uncertainty. Unlike traditional models that return a point estimate or a softmax probability, DREAM produces a patient-specific posterior that supports credible intervals and risk quantiles—enabling clinicians to move beyond thresholded decisions and toward spectrum-based, uncertainty-aware triage strategies in critical care.

The output is a posterior distribution on the mortality risk of 28 days, characterized by its mean, standard deviation, and credible interval (for example, 2.5th and 97.5th percentiles). Unlike deterministic outputs that may suffer from overconfidence or rigid thresholding, this distributional view enables clinicians to reason across a full spectrum of risk. For example, patients with moderate predicted mortality but wide uncertainty may prompt increased surveillance, while narrow intervals around high-risk estimates support confident triage decisions.

This approach allows individualized decision making in the ICU, aligning machine learning outputs with the probabilistic nature of clinical deterioration. By integrating posterior inference into the modeling framework, this study contributes to a robust and interpretable strategy for the evaluation of mortality risk in elderly patients with diabetes and heart failure.

\subsection*{Statistical Analyses and Interpretability Assessment}

To ensure the robustness, interpretability, and clinical applicability of our predictive framework, we implemented a structured suite of statistical and model explanation techniques, each serving a distinct yet complementary role in validating model integrity and aligning predictions with clinical reasoning.

We began by evaluating the comparability between the training and test cohorts through descriptive statistics and inferential testing. Specifically, two-sided t-tests (with Welch’s correction when variances were unequal) were applied to key continuous variables such as age, blood pressure, and laboratory markers. This step was essential to confirm that the dataset split preserved the underlying clinical distribution, thereby reducing sampling bias and enhancing the generalizability of performance metrics. Clinically, this ensures that model outputs are not driven by artifacts of population imbalance but reflect true signal reproducibility across different patient subgroups.

Next, we conducted an ablation analysis to systematically assess the marginal contribution of each selected feature. By retraining the model after the exclusion of each variable and measuring the resulting decrease in AUROC, we quantified its unique impact on predictive performance. This method offered a transparent mechanism for validating the necessity of retained features. From a clinical standpoint, it further elucidated which individual biomarkers carry the most prognostic weight, thus reinforcing the biological plausibility of the model’s core drivers.

Collectively, these statistical and interpretability strategies form a coherent validation framework that advances beyond performance benchmarking. By integrating population-level comparison, feature-level attribution, case-level transparency, and probabilistic uncertainty, our approach ensures that the proposed model is not only high-performing, but also trustworthy and clinically actionable. Notably, the incorporation of posterior predictive analysis represents a methodological innovation that bridges algorithmic prediction with personalized ICU care, and constitutes a core contribution of this study.

\section*{Results}
\subsection*{Cohort Characteristics and Statistical Comparison}

To assess the robustness and clinical reliability of the mortality prediction model in elderly patients with diabetes mellitus admitted to the ICU for heart failure, a comparative analysis was performed between the training and test cohorts, as well as between survivors and non-survivors. Table~\ref{tab:cohort comparison results} shows the distribution of 19 clinical features across the training (n = 1034) and test (n = 444) sets. Independent sample t-tests were conducted for each variable, and no statistically significant differences were observed in the majority of features. For instance, the average APS III score was 52.78 in the training set and 53.53 in the test set ($p = 0.490$), while the mean oxygen flow rates were also similar (7.46 vs 7.43 L/min, $p = 0.960$). These findings suggest that the training and test populations were well matched, which supports the model's ability to generalize to unseen cases and mitigates the risk of sampling bias.

Table~\ref{tab:cohort comparison results 1} presents comparisons between patients who survived beyond 28 days and those who died. A number of physiological and biochemical indicators showed statistically significant differences between the two groups. Notably, deceased patients had substantially higher APS III scores (64.28 vs 49.98, $p < 0.001$), as well as elevated levels of oxygen flow, anion gap, INR(PT), RDW-SD, phosphorus, and total bilirubin. These findings are consistent with multi-organ dysfunction and higher disease severity at ICU admission. In contrast, survivors had significantly higher GCS Eye Opening scores (3.30 vs 2.75, $p < 0.001$), and better Braden scale subcomponent scores such as mobility and nutrition, indicating reduced frailty and pressure injury risk.

Additionally, several binary variables demonstrated notable differences. The use of vasopressin and lorazepam was more frequent among patients who died (52\% and 58\%, respectively), which may reflect more aggressive interventions or underlying disease burden. The prevalence of severe sepsis with septic shock was also markedly higher among non-survivors (40\% vs 15\%, $p < 0.001$), underscoring its prognostic significance in this cohort. Differences in procedural variables such as transthoracic echocardiography and invasive ventilation also emerged, possibly indicating differing clinical trajectories or monitoring needs.

Altogether, the comparative analysis confirms both the clinical plausibility and predictive relevance of the selected features. The balance between training and test cohorts reinforces the model's external validity, while the observed stratification by survival status offers insight into the pathophysiological processes driving mortality. These results provide a strong foundation for the development of interpretable and generalizable predictive models in elderly ICU patients with diabetes and heart failure.

\begin{table}[H]
\noindent
\caption{\textbf{T-test Comparison of Feature Distributions between Training and Test Sets.}}
\label{tab:cohort comparison results}
\small
\renewcommand{\arraystretch}{1.2}
\rowcolors{2}{white}{white}
\begin{tabularx}{\textwidth}{>{\raggedright\arraybackslash}X|c|X|X|X}
\hline
\rowcolor[HTML]{D9EAD3}
\textbf{Feature} & \textbf{Unit} & \textbf{Training Set} & \textbf{Test Set} & \textbf{P-value} \\ \hline
apsiii & Score & 52.78 (18.31) & 53.53 (19.40) & 0.490 \\ \hline
GCS - Eye Opening & Score & 3.19 (0.85) & 3.16 (0.88) & 0.473 \\ \hline
O2 Flow & L/min & 7.46 (8.59) & 7.43 (8.71) & 0.960 \\ \hline
Braden Mobility & Score & 2.44 (0.54) & 2.46 (0.55) & 0.525 \\ \hline
INR(PT) & Ratio & 1.59 (0.73) & 1.54 (0.64) & 0.150 \\ \hline
Braden Nutrition & Score & 2.29 (0.45) & 2.31 (0.49) & 0.357 \\ \hline
RDW-SD & fL & 52.15 (8.06) & 51.71 (7.26) & 0.299 \\ \hline
pO2 & mmHg & 114.24 (64.13) & 108.45 (60.76) & 0.099 \\ \hline
Anion gap & mEq/L & 14.94 (4.02) & 15.19 (3.73) & 0.244 \\ \hline
age & Years & 75.79 (7.07) & 75.68 (6.93) & 0.776 \\ \hline
Phosphorous & mg/dL & 4.15 (1.21) & 4.21 (1.29) & 0.378 \\ \hline
Total Bilirubin & mg/dL & 0.99 (1.53) & 0.88 (0.76) & 0.085 \\ \hline
Vasopressin & Presence & 0.25 (0.43) & 0.25 (0.44) & 0.719 \\ \hline
Braden Friction/Shear & Score & 2.09 (0.40) & 2.10 (0.38) & 0.509 \\ \hline
LORazepam & Presence & 0.34 (0.48) & 0.42 (0.49) & 0.004 \\ \hline
Severe sepsis with septic shock & Presence & 0.20 (0.40) & 0.23 (0.42) & 0.181 \\ \hline
Multi Lumen & Presence & 0.26 (0.44) & 0.27 (0.44) & 0.784 \\ \hline
Transthoracic Echo & Presence & 0.18 (0.39) & 0.24 (0.43) & 0.025 \\ \hline
Invasive Ventilation & Presence & 0.39 (0.49) & 0.42 (0.49) & 0.303 \\ \hline
\end{tabularx}
\begin{flushleft}
\textbf{Table notes}: Summary statistics of key clinical variables for the training and test cohorts are presented as mean (SD). Between-group differences were assessed using independent t-tests, with statistical significance defined at a threshold of $p < 0.05$.
\end{flushleft}
\end{table}

\begin{table}[H]
\noindent
\caption{\textbf{T-test Comparison of Feature Distributions between Survival and Non-Survival Sets.}}
\label{tab:cohort comparison results 1}
\small
\renewcommand{\arraystretch}{1.2}
\rowcolors{2}{white}{white}
\begin{tabularx}{\textwidth}{>{\raggedright\arraybackslash}X|c|X|X|X}
\hline
\rowcolor[HTML]{D9EAD3}
\textbf{Feature} & \textbf{Unit} & \textbf{Survival} & \textbf{Non-Survival} & \textbf{P-value} \\ \hline
apsiii & Score & 49.98 (16.65) & 64.28 (20.24) & $<$ 0.001 \\ \hline
GCS - Eye Opening & Score & 3.30 (0.75) & 2.75 (1.04) & $<$ 0.001 \\ \hline
O2 Flow & L/min & 6.66 (7.32) & 10.71 (12.01) & $<$ 0.001 \\ \hline
Braden Mobility & Score & 2.51 (0.51) & 2.14 (0.54) & $<$ 0.001 \\ \hline
INR(PT) & Ratio & 1.52 (0.66) & 1.86 (0.92) & $<$ 0.001 \\ \hline
Braden Nutrition & Score & 2.35 (0.44) & 2.03 (0.38) & $<$ 0.001 \\ \hline
RDW-SD & fL & 51.50 (7.59) & 54.82 (9.34) & $<$ 0.001 \\ \hline
pO2 & mmHg & 118.11 (66.41) & 98.40 (50.99) & $<$ 0.001 \\ \hline
Anion gap & mEq/L & 14.51 (3.73) & 16.70 (4.68) & $<$ 0.001 \\ \hline
age & Years & 75.43 (7.07) & 77.26 (6.89) & $<$ 0.001 \\ \hline
Phosphorous & mg/dL & 4.06 (1.14) & 4.50 (1.39) & $<$ 0.001 \\ \hline
Total Bilirubin & mg/dL & 0.89 (1.13) & 1.39 (2.55) & 0.006 \\ \hline
Vasopressin & Presence & 0.18 (0.38) & 0.52 (0.50) & $<$ 0.001 \\ \hline
Braden Friction/Shear & Score & 2.14 (0.40) & 1.90 (0.38) & $<$ 0.001 \\ \hline
LORazepam & Presence & 0.29 (0.45) & 0.58 (0.50) & $<$ 0.001 \\ \hline
Severe sepsis with septic shock & Presence & 0.15 (0.36) & 0.40 (0.49) & $<$ 0.001 \\ \hline
Multi Lumen & Presence & 0.23 (0.42) & 0.39 (0.49) & $<$ 0.001 \\ \hline
Transthoracic Echo & Presence & 0.17 (0.37) & 0.26 (0.44) & 0.007 \\ \hline
Invasive Ventilation & Presence & 0.38 (0.48) & 0.47 (0.50) & 0.013 \\ \hline
\end{tabularx}
\begin{flushleft}
\textbf{Table notes}: This table compares patients who survived versus those who died within 28 days. Differences in mean values of key clinical variables are reported alongside corresponding $p$-values. Statistical significance was defined at a threshold of $p < 0.05$.
\end{flushleft}
\end{table}

\subsection*{Ablation Analysis}

\begin{figure}[H]
    \centering
    \includegraphics[width=0.95\linewidth]{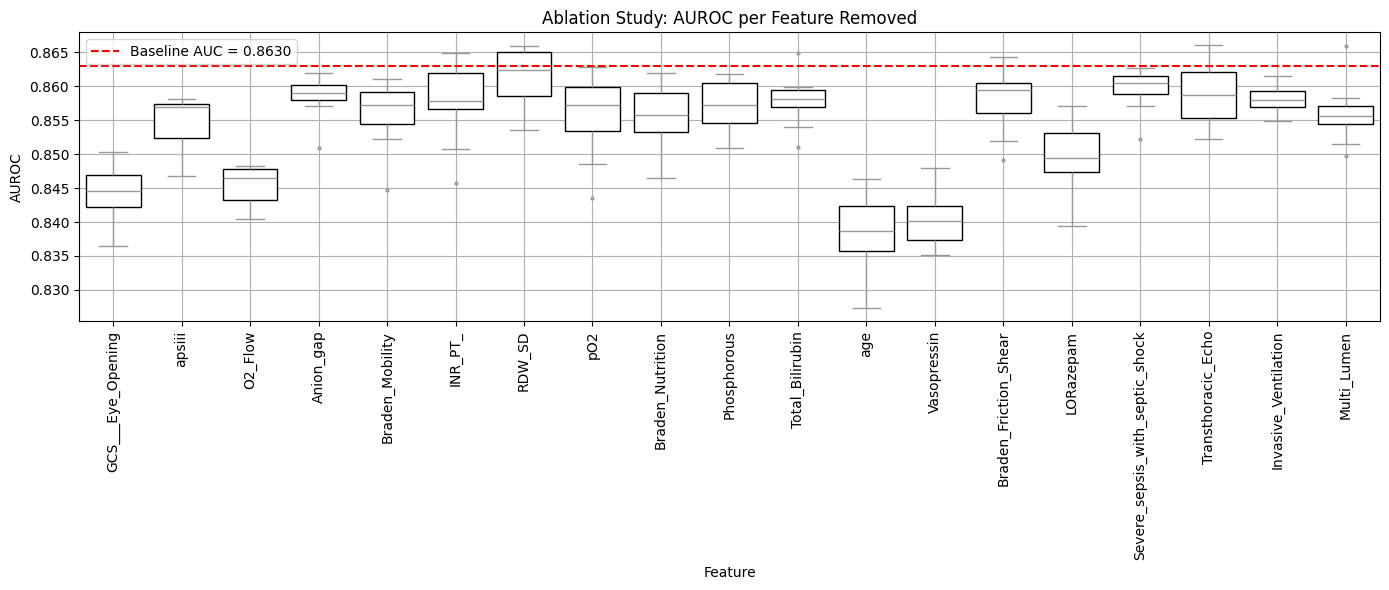}
    \caption{\textbf{Impact of Feature Removal on LR Model Performance.}}
    \label{fig:ablation analysis}
\end{figure}

To evaluate the robustness and clinical interpretability of the logistic regression model in predicting 28-day mortality among elderly ICU patients with diabetes mellitus and heart failure, we conducted an ablation study. As illustrated in Figure~\ref{fig:ablation analysis}, each predictor was individually removed from the model input, followed by model retraining and performance evaluation using bootstrap sampling. The impact of each variable was quantified based on its effect on the AUROC metric. The red dashed line in the figure denotes the baseline AUROC of 0.8630 achieved by the full model incorporating all selected features.

The ablation results revealed that the removal of several variables caused a noticeable drop in model performance. In particular, the absence of age significantly reduced AUROC, highlighting its critical role in risk stratification among older patients in the ICU. Likewise, the removal of vasopressin administration and the GCS eye opening score led to a marked decrease in predictive performance, suggesting that both hemodynamic instability and neurologic responsiveness contribute substantially to the risk of early mortality. 

However, variables such as RDW(SD) and transthoracic echo had a limited impact on AUROC when omitted, indicating a relatively modest marginal effect on model discrimination. Although these variables contribute little to the ablation of logistic regression, tree-based models such as CatBoost and LightGBM can exploit their non-linear effects and interactions to enhance discrimination in mortality prediction.

The structure of the logistic regression model further enhances interpretability. Each coefficient reflects a direct relationship with the outcome, allowing clinicians to infer logical clinical reasoning from the statistical patterns. The observed alignment between ablation results and known prognostic indicators, such as the severity of the disease (APS III), oxygen support requirements, and coagulation abnormalities (INR[PT]), reinforces the biological plausibility of the model's decision pathway.

In summary, the ablation study supports both the robustness and the medical relevance of the logistic model in predicting short-term mortality in this high-risk elderly population. The consistent AUROC under perturbations and the interpretable behavior of key features suggest that such a model could be used for real-time triage and decision support in critical care.

\subsection*{Model Performance Evaluation and Clinical Interpretability}

\begin{figure}[H]
\centering
\includegraphics[width=0.95\linewidth]{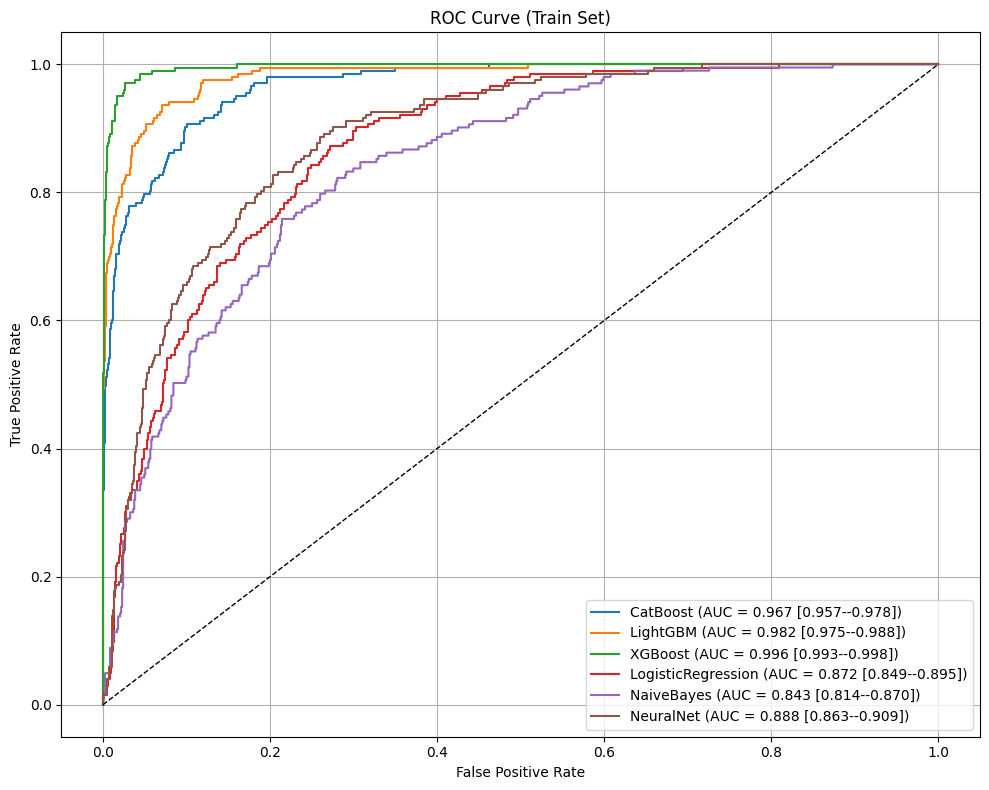}
\caption{\textbf{AUROC for Model Performance in the Training Set.}}
\label{fig:roc_train}
\end{figure}

\begin{figure}[H]
\centering
\includegraphics[width=0.95\linewidth]{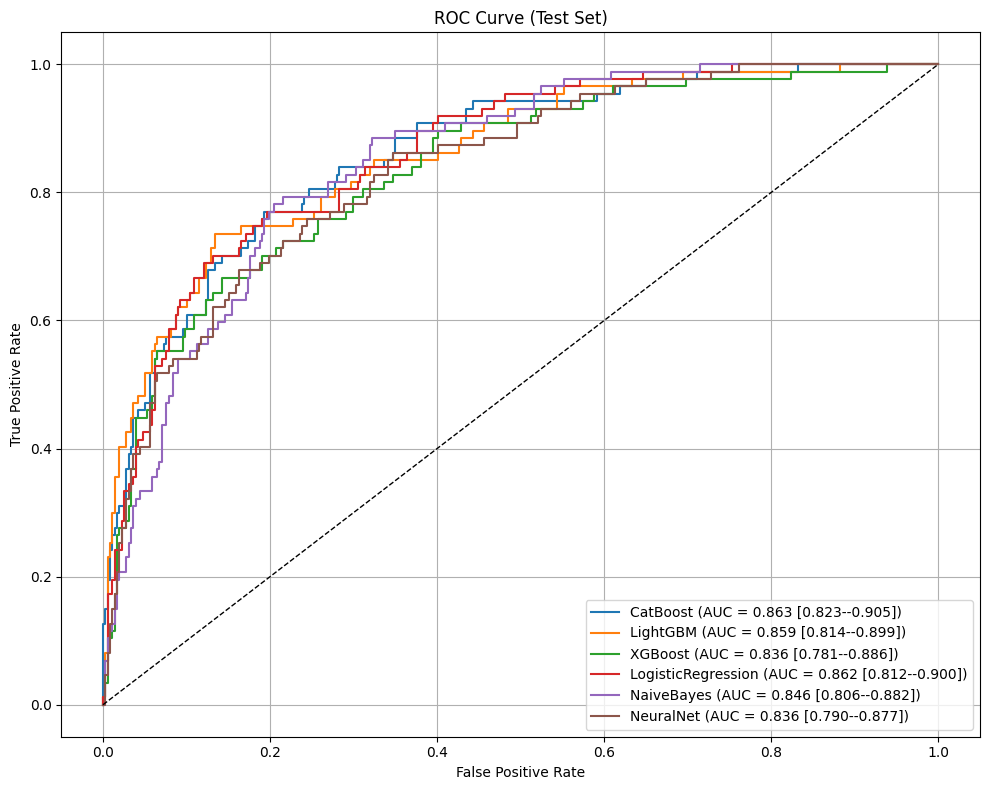}
\caption{\textbf{AUROC for Model Performance in the Test Set.}}
\label{fig:roc_test}
\end{figure}

To thoroughly evaluate the predictive performance and robustness of various machine learning models in estimating 28-day mortality among elderly ICU patients with diabetes mellitus and heart failure, we conducted a comparative analysis of six representative algorithms. Key evaluation metrics—including AUROC, accuracy, sensitivity, specificity, positive predictive value (PPV), negative predictive value (NPV), and F1 score—are summarized in Table~\ref{tab: Results of the Training Set} and Table~\ref{tab: Results of the Test Set}. The corresponding ROC curves for both the training and test cohorts are depicted in Figures~\ref{fig:roc_train} and~\ref{fig:roc_test}, respectively.

In the training set, all ensemble-based models achieved excellent results. XGBoost had the highest AUROC of 0.996 (95\% CI: 0.993--0.998), followed by LightGBM (0.982) and CatBoost (0.967), with all three demonstrating outstanding discrimination and sensitivity. Logistic regression, Naive Bayes, and neural network models yielded lower but acceptable performance, reflecting the limitations of simpler or more constrained learning algorithms in high-dimensional clinical data.

Evaluation of the test set provided information on model generalizability. CatBoost remained the leader with an AUROC of 0.863 (95\% CI: 0.823--0.905), closely followed by LightGBM (0.859) and logistic regression (0.862). We manually adjusted the threshold to ensure that the sensitivity was above 0.8, ensuring that the model could distinguish patients with high risk of death as much as possible while not misjudging low-risk groups (based on its relatively high specificity).

These results confirm the suitability of gradient-boosting models in handling the heterogeneity and missingness common in ICU datasets. CatBoost, in particular, performed robustly in both the training and testing phases, likely benefiting from its ability to natively process categorical variables and avoid overfitting through ordered boosting. In addition, its consistent prioritization of physiologically meaningful features, such as the APSIII score, oxygen demand, and nutritional / mobility status, adds interpretability and clinical relevance to its predictions.

In summary, CatBoost emerged as a reliable and interpretable model for predicting mortality in elderly ICU patients with diabetes mellitus and heart failure. Its favorable performance metrics, minimal performance degradation in data sets, and alignment with established clinical indicators highlight its promise for integration into ICU decision support workflows.

\begin{table}[H]
\small
\renewcommand{\arraystretch}{1.2}
\centering
\caption{\textbf{Performance Comparison of Different Models in the Training Set.}}
\begin{tabular}{l|l|l|l|l|l|l|l}
\hline
\rowcolor[HTML]{D9EAD3}
\textbf{Model} & \textbf{AUC (95\% CI)} & \textbf{Accuracy} & \textbf{F1-score} & \textbf{Sensitivity} & \textbf{Specificity} & \textbf{PPV} & \textbf{NPV} \\ \hline
\rowcolor[HTML]{FDE9D9}
CatBoost & 0.967 (0.957--0.978) & 0.920 & 0.793 & 0.783 & 0.953 & 0.803 & 0.947 \\ \hline
LightGBM & 0.982 (0.975--0.988) & 0.940 & 0.844 & 0.828 & 0.968 & 0.862 & 0.958 \\ \hline
XGBoost & 0.996 (0.993--0.998) & 0.975 & 0.936 & 0.941 & 0.983 & 0.932 & 0.986 \\ \hline
LR & 0.872 (0.849--0.895) & 0.787 & 0.585 & 0.764 & 0.793 & 0.474 & 0.932 \\ \hline
NaiveBayes & 0.843 (0.814--0.870) & 0.779 & 0.548 & 0.685 & 0.801 & 0.457 & 0.912 \\ \hline
NeuralNet & 0.888 (0.863--0.909) & 0.752 & 0.588 & 0.901 & 0.716 & 0.437 & 0.967 \\ \hline
\end{tabular}
\label{tab: Results of the Training Set}
\end{table}

\begin{table}[H]
\small
\renewcommand{\arraystretch}{1.2}
\centering
\caption{\textbf{Performance Comparison of Different Models in the Test Set.}}
\begin{tabular}{l|l|l|l|l|l|l|l}
\hline
\rowcolor[HTML]{D9EAD3}
\textbf{Model} & \textbf{AUC (95\% CI)} & \textbf{Accuracy} & \textbf{F1-score} & \textbf{Sensitivity} & \textbf{Specificity} & \textbf{PPV} & \textbf{NPV} \\ \hline
\rowcolor[HTML]{FDE9D9}
CatBoost & 0.863 (0.823--0.905) & 0.764 & 0.571 & 0.805 & 0.754 & 0.443 & 0.941 \\ \hline
LightGBM & 0.859 (0.814--0.899) & 0.739 & 0.547 & 0.805 & 0.723 & 0.414 & 0.938 \\ \hline
XGBoost & 0.836 (0.781--0.886) & 0.712 & 0.522 & 0.805 & 0.689 & 0.387 & 0.935 \\ \hline
LR & 0.862 (0.812--0.900) & 0.734 & 0.543 & 0.805 & 0.717 & 0.409 & 0.938 \\ \hline
NaiveBayes & 0.846 (0.806--0.882) & 0.748 & 0.559 & 0.816 & 0.731 & 0.425 & 0.942 \\ \hline
NeuralNet & 0.836 (0.790--0.877) & 0.707 & 0.522 & 0.816 & 0.681 & 0.384 & 0.938 \\ \hline
\end{tabular}
\label{tab: Results of the Test Set}
\end{table}

\subsection*{Posterior Prediction of Individual Mortality Risk}

To evaluate not only the robustness of the CatBoost model but also to generate individualized mortality risk distributions, we employed posterior inference via the DREAM algorithm. Unlike point estimates, this distributional output enables clinicians to assess uncertainty and stratify risk on a continuous spectrum, which is particularly valuable for nuanced decision-making in critical care.

The input priors were sampled from the empirical distributions of non-survivors detailed in Table~\ref{tab:cohort comparison results 1}, thus constructing a simulated high-risk patient profile grounded in real ICU outcomes. This ensures that the resulting probability distribution reflects clinically plausible conditions, rather than artificial or average scenarios.

\begin{figure}[H]
    \centering
    \includegraphics[width=0.85\linewidth]{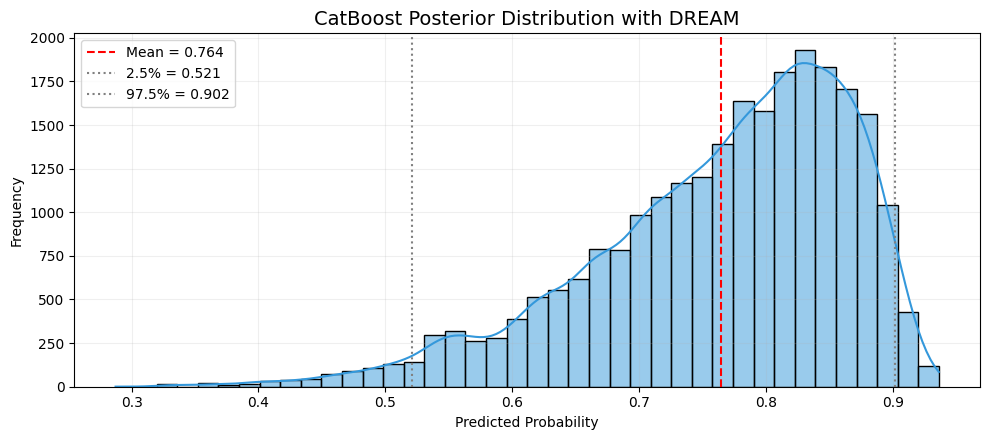}
    \caption{\textbf{Posterior distribution of 28-day mortality for a high-risk elderly diabetes mellitus \& heart failure ICU patient.}}
    \label{fig:uq_catboost}
\end{figure}

As shown in Figure~\ref{fig:uq_catboost}, the predicted risk distribution displays a right-skewed pattern, with a mean of 0.764, and 2.5\% and 97.5\% quantiles of 0.521 and 0.902, respectively. This credible interval reflects the model’s confidence when evaluating a patient whose physiology mirrors that of the actual non-survivor group. Such quantification is critical in ICU settings, where decisions often hinge on the presence of multiple overlapping risk signals.

The simulated input configuration included APSIII = 64.3, age = 77.3, O2 flow = 10.7, INR(PT) = 1.86 and Braden mobility = 2.14 - variables that all showed significant separation in Table~\ref{tab:cohort comparison results 1}. The presence of GCS eye = 2.75 and Braden flexion = 1.90 further reinforced the poor functional and neurological status of the profile. These combinations result in consistently elevated predicted risk, suggesting that the model correctly synthesizes critical factors that drive mortality in elderly patients with diabetes mellitus and heart failure.

When compared to the mean probability of death in our sample (0.196, equivalent to a random guess), the 2.5\% quantile of the posterior mortality risk for the high-risk group was 0.521, which is substantially higher. This indicates that the model is capable of effective stratification. In clinical practice, the DREAM-based framework can be used to assess individualized mortality risk. By specifying patient-level parameters in the \texttt{feature\_priors} variable, the predicted 28-day mortality probability can be derived, facilitating interpretable and patient-specific decision-making.

Importantly, this analysis highlights the ability of the model to provide individualized risk estimates that are both probabilistically meaningful and clinically relevant. Rather than generating a static score, the model outputs a full posterior distribution conditioned on population-informed uncertainty. For decision makers, this allows risk communication to move beyond binary classification into a quantified confidence framework, especially valuable in high-stakes environments such as the ICU.

\subsection*{ALE Analysis and Clinical Interpretability}

\begin{figure}[H]
    \centering
    \includegraphics[width=0.85\linewidth]{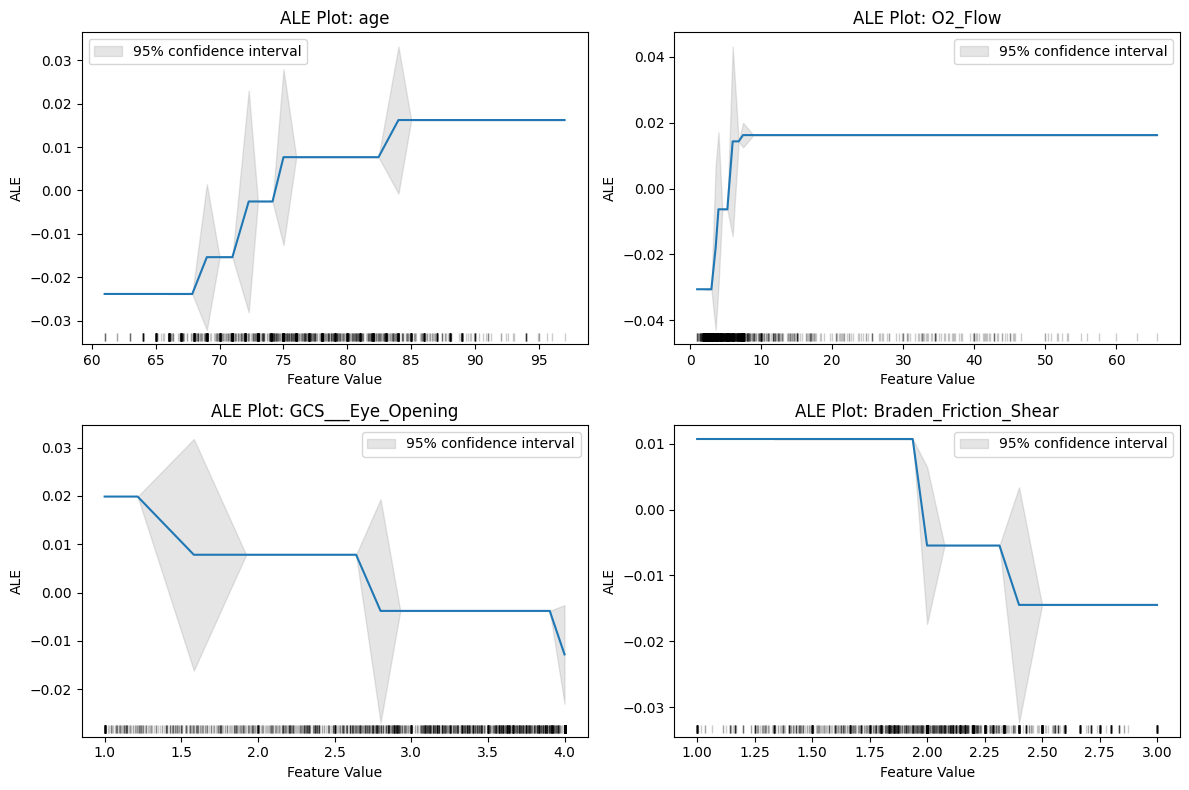}
    \caption{\textbf{ALE plots for top features in elderly ICU patients with diabetes and heart failure.}}
    \label{fig:ale_analysis}
\end{figure}

To enhance the interpretability of the predictive model developed for 28-day mortality in elderly ICU patients with diabetes mellitus and heart failure, we conducted ALE analysis on selected high-impact features. Figure~\ref{fig:ale_analysis} presents ALE plots for four representative variables: age, O2 Flow, GCS Eye Opening, and Braden Friction/Shear.

The ALE plot of age reveals a modest but consistent upward trend in predicted mortality risk beginning around 70 years, with the effect becoming more pronounced beyond 80. This corresponds well with the age distribution in the cohort and reflects the well-established association between advanced age and worse ICU outcomes. The effect appears linear and stable, indicating the model's ability to capture age-related vulnerability without abrupt or unrealistic risk shifts.

In the case of O2 Flow, the ALE curve shows a sharp increase in risk between 0 and 8 L/min, with a plateau afterward. The density of observations is concentrated below 10 L/min, suggesting that small increases in oxygen requirements within this range may be early markers of respiratory deterioration. Beyond this threshold, the ALE curve flattens, implying that extremely high flow rates are less informative, possibly due to saturation effects or reduced sample support.

For the GCS Eye Opening component, the ALE function indicates a stepwise decline in predicted risk as the score improves from 1 to 4. Patients with a score of 1 exhibited a notable positive shift in mortality risk, whereas those with a score of 4 approached a neutral effect. This aligns with clinical understanding that lower GCS scores reflect impaired consciousness and higher acuity. The distribution of data points shows sufficient support across all levels, enhancing confidence in this interpretation.

The Braden Friction/Shear score also demonstrates clear interpretability: patients with lower scores (closer to 1) are associated with increased mortality risk, likely due to greater immobility and risk of pressure injury. The ALE curve drops markedly between scores of 2.0 and 2.5, stabilizing thereafter. The presence of a consistent negative gradient across this narrow score range reflects the sensitivity of the model to subtle indicators of frailty.

In general, these ALE results support the robustness of the model and its alignment with pathophysiological mechanisms relevant to ICU outcomes in elderly patients with diabetes mellitus and heart failure. Importantly, the curves are smooth, monotonic in clinically expected directions, and underpinned by well-distributed feature values. These findings reinforce the validity of the model, both in terms of predictive performance and biological plausibility, which is critical for clinical deployment and user trust in decision support tools.

\section{Discussion}\label{sec4}
\subsection{Summary of Existing Model Compilation}
In this study, we developed a machine learning model to predict short-term mortality in elderly ICU patients (65–90 years) with coexisting diabetes mellitus and heart failure—two comorbidities that jointly elevate risk through overlapping mechanisms such as insulin resistance and impaired organ function. Unlike prior models focused on binary classification, our framework produces full mortality risk distributions. This distributional output offers clinicians a clearer picture of uncertainty and supports more nuanced decision-making, especially for borderline or high-risk cases. Despite the clinical importance of this population, few existing tools offer this level of individualized, probabilistic insight.

To address this clinical gap, we constructed a tailored machine learning framework using the MIMIC-IV database, focusing on 1,478 elderly ICU patients. The model development pipeline was designed to be rigorous and reproducible. We first excluded individuals outside the target age range or with active malignancy to define a more clinically homogeneous cohort. From the initial dataset, over 600 variables were extracted across six core domains: demographics, laboratory results, charted assessments, medication records, procedures, and comorbidities.

To refine these features, we applied a two-stage selection process. The first stage used ANOVA F-statistics to filter top-performing variables, and the second employed a Random Forest–based ranking to assess their relative importance. Ultimately, 19 clinically interpretable predictors were selected. These included indicators of physiological instability (e.g., anion gap, Braden subscale scores), treatment intensity (e.g., invasive ventilation), and organ dysfunction (e.g., INR(PT), pO\textsubscript{2}, APS III score).

Among six algorithms tested, CatBoost demonstrated the best performance, with a test AUROC exceeding 0.86 and balanced sensitivity and specificity. Unlike traditional scoring systems, this approach accommodated complex nonlinear interactions and missingness common in ICU data. Importantly, SMOTE resampling during cross-validation preserved generalizability in imbalanced datasets, while DREAM-based posterior inference provided individual-level risk estimates with uncertainty quantification—an important clinical advantage.

This model offers critical clinical value: (1) it supports early identification of elderly diabetic heart failure patients at high risk of in-hospital death, (2) it allows targeted allocation of ICU resources (e.g. palliative consultation, intensified monitoring), and (3) it enhances shared decision-making by providing interpretable probabilistic risk outputs. The inclusion of frailty markers such as Braden Mobility and Nutrition scores further emphasizes its relevance to geriatric critical care, where standard severity scores may lack sensitivity.

\subsection{Comparison with Prior Studies}
While prior studies have examined mortality risks from diabetes or heart failure independently, few have addressed their combined impact in elderly ICU patients. The complexity of managing both diseases simultaneously—especially in critically ill older adults—remains underexplored. Moreover, most existing models offer only point or binary predictions, lacking uncertainty quantification. In contrast, our model provides full posterior risk distributions, enabling individualized and probabilistically informed ICU decision-making.

A study by Aguilar et al. (2009) observed that glycemic control levels were associated with differences in long-term survival among diabetic patients with heart failure. They noted a U-shaped relationship between HbA1C levels and mortality, with moderate control linked to the best outcomes. However, their findings were based on outpatient data from a Veterans Affairs cohort and relied on conventional regression techniques, making it difficult to apply their conclusions directly to ICU patients facing acute complications.

Evans et al. (2010) investigated the use of metformin in individuals with both diabetes and chronic heart failure, concluding that metformin treatment was associated with improved survival. While important, their research focused on a broader adult population and did not consider ICU admissions or elderly-specific patterns of deterioration.

More recently, Zhou et al. (2023) introduced the stress hyperglycemia ratio (SHR) as a prognostic marker in patients with diabetes and acute decompensated heart failure. Their results also revealed a U-shaped association between SHR and long-term mortality. Although their study addressed hospitalization and used modern statistical modeling, it did not focus on ICU mortality, and SHR alone may not capture the full clinical picture during early critical illness.

Compared to these previous works, our study focuses on a distinctly vulnerable population: elderly ICU patients with both diabetes and heart failure. Using a comprehensive set of features available within the first 24 hours of admission to the ICU and applying advanced machine learning techniques, our aim was to improve early mortality prediction in a high-risk real-world setting. Our approach integrates clinical, physiological, and treatment-related variables, allowing us to model complex interactions that traditional methods may overlook. In doing so, we fill a significant gap in the literature, offering a more practical and tailored risk stratification tool for geriatric critical care.

\subsection{Limitations and Future Work}
Despite encouraging results, this study has several notable limitations. First, although the MIMIC-IV database provides detailed and well-curated ICU data, it represents a single healthcare system, potentially limiting the model’s external generalizability. External validation using multi-center data—particularly from non-U.S. institutions or community hospitals—is essential before clinical deployment.

Second, the use of structured data excludes unstructured clinical notes, radiographic findings, and genomic information that can further improve prognostic precision. Incorporating natural language processing (NLP) techniques to extract insights from clinical narratives or integrating radiology reports could improve prediction.

Third, while our model includes ablation-based interpretability and posterior distribution inference, additional transparency could be achieved by integrating SHAP or LIME explanations for bedside decision support. These tools can provide clinicians with a clearer rationale for model predictions, enhancing clinical trust and adoption.

Finally, future research should explore dynamic prediction models that continuously update mortality risk as new data becomes available during the ICU stay. Real-time mortality forecasting could better reflect the evolving clinical course and support more responsive care planning.

\section{Conclusion}\label{sec5}
We conducted a focused analysis on elderly ICU patients diagnosed with both diabetes and heart failure—two conditions that often coexist and significantly impact short-term survival in critical care. Drawing from the MIMIC-IV database, we built and evaluated a predictive model that uses variables collected within the first 24 hours of ICU admission. This approach allows clinicians to identify high-risk patients early and more accurately than traditional risk scoring systems.

Among several machine learning methods tested, the CatBoost model delivered the strongest results, showing high discriminative power and robustness across patient subgroups. By incorporating a variety of clinical features, such as lab results, chart events, comorbidities, and procedures, our model goes beyond conventional tools to reflect the complexity of real-world ICU decision making.

Unlike earlier studies that often focused on general hospital or outpatient cohorts, our work is one of the few that targets elderly patients with both diabetes and heart failure in the ICU setting. The strength of the model lies not only in its precision, but also in its ability to provide interpretable, individualized risk estimates, which is crucial in time-sensitive and resource-constrained clinical environments.

Looking ahead, validating this model in external hospital systems, integrating unstructured data like clinician notes, and expanding to include long-term outcomes could make this tool even more practical. Ultimately, this study lays important groundwork for building smarter, more adaptive decision support systems tailored to aging populations with complex health needs.

\section*{Declarations}

\subsection*{Funding}

Not applicable.

\subsection*{Conflict of interest/Competing interests}

The authors declare that the research was conducted in the absence of any commercial or financial relationships that could be construed as a potential conflict of interest.

\subsection*{Ethics approval and consent to participate}

This study used deidentified, publicly available databases (MIMIC-IV and eICU Collaborative Research Database) and does not require institutional review board (IRB) approval or patient consent.

\subsection*{Consent for publication}

Not applicable.

\subsection*{Data availability}

The datasets analyzed in this study are publicly available from PhysioNet:

- MIMIC-IV: \url{https://physionet.org/content/mimiciv/}

\subsection*{Materials availability}

Not applicable.

\subsection*{Code availability}

Not applicable.

\subsection*{Clinical Trial Number}

Not applicable.

\subsection*{Author contributions}

JF was responsible for the conception of the study and oversaw the design and execution of all experiments. SC, LS and YS jointly handled data pre-processing, model development, and manuscript writing. EP, KA, GP and MP contributed medical expertise, offering clinical interpretation and ensuring alignment of results with patient care practices. MP had overall responsibility for project supervision. All authors contributed to the final manuscript and approved its submission.

\bibliography{sn-bibliography}


\begin{thebibliography}{40}
\ifx \bisbn   \undefined \def \bisbn  #1{ISBN #1}\fi
\ifx \binits  \undefined \def \binits#1{#1}\fi
\ifx \bauthor  \undefined \def \bauthor#1{#1}\fi
\ifx \batitle  \undefined \def \batitle#1{#1}\fi
\ifx \bjtitle  \undefined \def \bjtitle#1{#1}\fi
\ifx \bvolume  \undefined \def \bvolume#1{\textbf{#1}}\fi
\ifx \byear  \undefined \def \byear#1{#1}\fi
\ifx \bissue  \undefined \def \bissue#1{#1}\fi
\ifx \bfpage  \undefined \def \bfpage#1{#1}\fi
\ifx \blpage  \undefined \def \blpage #1{#1}\fi
\ifx \burl  \undefined \def \burl#1{\textsf{#1}}\fi
\ifx \doiurl  \undefined \def \doiurl#1{\url{https://doi.org/#1}}\fi
\ifx \betal  \undefined \def \betal{\textit{et al.}}\fi
\ifx \binstitute  \undefined \def \binstitute#1{#1}\fi
\ifx \binstitutionaled  \undefined \def \binstitutionaled#1{#1}\fi
\ifx \bctitle  \undefined \def \bctitle#1{#1}\fi
\ifx \beditor  \undefined \def \beditor#1{#1}\fi
\ifx \bpublisher  \undefined \def \bpublisher#1{#1}\fi
\ifx \bbtitle  \undefined \def \bbtitle#1{#1}\fi
\ifx \bedition  \undefined \def \bedition#1{#1}\fi
\ifx \bseriesno  \undefined \def \bseriesno#1{#1}\fi
\ifx \blocation  \undefined \def \blocation#1{#1}\fi
\ifx \bsertitle  \undefined \def \bsertitle#1{#1}\fi
\ifx \bsnm \undefined \def \bsnm#1{#1}\fi
\ifx \bsuffix \undefined \def \bsuffix#1{#1}\fi
\ifx \bparticle \undefined \def \bparticle#1{#1}\fi
\ifx \barticle \undefined \def \barticle#1{#1}\fi
\bibcommenthead
\ifx \bconfdate \undefined \def \bconfdate #1{#1}\fi
\ifx \botherref \undefined \def \botherref #1{#1}\fi
\ifx \url \undefined \def \url#1{\textsf{#1}}\fi
\ifx \bchapter \undefined \def \bchapter#1{#1}\fi
\ifx \bbook \undefined \def \bbook#1{#1}\fi
\ifx \bcomment \undefined \def \bcomment#1{#1}\fi
\ifx \oauthor \undefined \def \oauthor#1{#1}\fi
\ifx \citeauthoryear \undefined \def \citeauthoryear#1{#1}\fi
\ifx \endbibitem  \undefined \def \endbibitem {}\fi
\ifx \bconflocation  \undefined \def \bconflocation#1{#1}\fi
\ifx \arxivurl  \undefined \def \arxivurl#1{\textsf{#1}}\fi
\csname PreBibitemsHook\endcsname

\bibitem[\protect\citeauthoryear{Foster}{1915}]{foster1915diabetes}
\begin{bbook}
\bauthor{\bsnm{Foster}, \binits{N.B.}}:
\bbtitle{Diabetes Mellitus}.
\bpublisher{Lippincott}, \blocation{???}
(\byear{1915})
\end{bbook}
\endbibitem

\bibitem[\protect\citeauthoryear{Alam et~al.}{2014}]{alam2014general}
\begin{barticle}
\bauthor{\bsnm{Alam}, \binits{U.}},
\bauthor{\bsnm{Asghar}, \binits{O.}},
\bauthor{\bsnm{Azmi}, \binits{S.}},
\bauthor{\bsnm{Malik}, \binits{R.A.}}:
\batitle{General aspects of diabetes mellitus}.
\bjtitle{Handbook of clinical neurology}
\bvolume{126},
\bfpage{211}--\blpage{222}
(\byear{2014})
\end{barticle}
\endbibitem

\bibitem[\protect\citeauthoryear{Kaul et~al.}{2013}]{kaul2013introduction}
\begin{botherref}
\oauthor{\bsnm{Kaul}, \binits{K.}},
\oauthor{\bsnm{Tarr}, \binits{J.M.}},
\oauthor{\bsnm{Ahmad}, \binits{S.I.}},
\oauthor{\bsnm{Kohner}, \binits{E.M.}},
\oauthor{\bsnm{Chibber}, \binits{R.}}:
Introduction to diabetes mellitus.
Diabetes: an old disease, a new insight,
1--11
(2013)
\end{botherref}
\endbibitem

\bibitem[\protect\citeauthoryear{Blair}{2016}]{blair2016diabetes}
\begin{botherref}
\oauthor{\bsnm{Blair}, \binits{M.}}:
Diabetes mellitus review.
Urologic nursing
\textbf{36}(1)
(2016)
\end{botherref}
\endbibitem

\bibitem[\protect\citeauthoryear{{Centers for Disease Control and Prevention}}{2023}]{cdc2023}
\begin{botherref}
\oauthor{\bsnm{{Centers for Disease Control and Prevention}}}:
National Diabetes Statistics Report: Estimates of Diabetes and Its Burden in the United States.
U.S. Department of Health and Human Services
(2023).
\url{https://www.cdc.gov/diabetes/php/data-research/index.html}
\end{botherref}
\endbibitem

\bibitem[\protect\citeauthoryear{Federation}{2025}]{idf2025}
\begin{botherref}
\oauthor{\bsnm{Federation}, \binits{I.D.}}:
IDF Diabetes Atlas (11th Edition).
\url{https://diabetesatlas.org/resources/idf-diabetes-atlas-2025/}.
Accessed: 2025-05-27
(2025)
\end{botherref}
\endbibitem

\bibitem[\protect\citeauthoryear{Papatheodorou et~al.}{2018}]{papatheodorou2018complications}
\begin{barticle}
\bauthor{\bsnm{Papatheodorou}, \binits{K.}},
\bauthor{\bsnm{Banach}, \binits{M.}},
\bauthor{\bsnm{Bekiari}, \binits{E.}},
\bauthor{\bsnm{Rizzo}, \binits{M.}},
\bauthor{\bsnm{Edmonds}, \binits{M.}}:
\batitle{Complications of diabetes 2017}.
\bjtitle{Journal of diabetes research}
\bvolume{2018},
\bfpage{3086167}
(\byear{2018})
\end{barticle}
\endbibitem

\bibitem[\protect\citeauthoryear{Gregg et~al.}{2016}]{gregg2016changing}
\begin{barticle}
\bauthor{\bsnm{Gregg}, \binits{E.W.}},
\bauthor{\bsnm{Sattar}, \binits{N.}},
\bauthor{\bsnm{Ali}, \binits{M.K.}}:
\batitle{The changing face of diabetes complications}.
\bjtitle{The lancet Diabetes \& endocrinology}
\bvolume{4}(\bissue{6}),
\bfpage{537}--\blpage{547}
(\byear{2016})
\end{barticle}
\endbibitem

\bibitem[\protect\citeauthoryear{Kannel and McGee}{1979}]{kannel1979diabetes}
\begin{barticle}
\bauthor{\bsnm{Kannel}, \binits{W.B.}},
\bauthor{\bsnm{McGee}, \binits{D.L.}}:
\batitle{Diabetes and cardiovascular disease: the framingham study}.
\bjtitle{Jama}
\bvolume{241}(\bissue{19}),
\bfpage{2035}--\blpage{2038}
(\byear{1979})
\end{barticle}
\endbibitem

\bibitem[\protect\citeauthoryear{Bertoni et~al.}{2004}]{bertoni2004heart}
\begin{barticle}
\bauthor{\bsnm{Bertoni}, \binits{A.G.}},
\bauthor{\bsnm{Hundley}, \binits{W.G.}},
\bauthor{\bsnm{Massing}, \binits{M.W.}},
\bauthor{\bsnm{Bonds}, \binits{D.E.}},
\bauthor{\bsnm{Burke}, \binits{G.L.}},
\bauthor{\bsnm{Goff~Jr}, \binits{D.C.}}:
\batitle{Heart failure prevalence, incidence, and mortality in the elderly with diabetes}.
\bjtitle{Diabetes care}
\bvolume{27}(\bissue{3}),
\bfpage{699}--\blpage{703}
(\byear{2004})
\end{barticle}
\endbibitem

\bibitem[\protect\citeauthoryear{Verny}{2007}]{verny2007congestive}
\begin{barticle}
\bauthor{\bsnm{Verny}, \binits{C.}}:
\batitle{Congestive heart failure in the elderly diabetic}.
\bjtitle{Diabetes \& metabolism}
\bvolume{33},
\bfpage{32}--\blpage{39}
(\byear{2007})
\end{barticle}
\endbibitem

\bibitem[\protect\citeauthoryear{MacDonald et~al.}{2008}]{macdonald2008diabetes}
\begin{barticle}
\bauthor{\bsnm{MacDonald}, \binits{M.R.}},
\bauthor{\bsnm{Petrie}, \binits{M.C.}},
\bauthor{\bsnm{Hawkins}, \binits{N.M.}},
\bauthor{\bsnm{Petrie}, \binits{J.R.}},
\bauthor{\bsnm{Fisher}, \binits{M.}},
\bauthor{\bsnm{McKelvie}, \binits{R.}},
\bauthor{\bsnm{Aguilar}, \binits{D.}},
\bauthor{\bsnm{Krum}, \binits{H.}},
\bauthor{\bsnm{McMurray}, \binits{J.J.}}:
\batitle{Diabetes, left ventricular systolic dysfunction, and chronic heart failure}.
\bjtitle{European heart journal}
\bvolume{29}(\bissue{10}),
\bfpage{1224}--\blpage{1240}
(\byear{2008})
\end{barticle}
\endbibitem

\bibitem[\protect\citeauthoryear{Shah et~al.}{2017}]{shah2017heart}
\begin{barticle}
\bauthor{\bsnm{Shah}, \binits{K.S.}},
\bauthor{\bsnm{Xu}, \binits{H.}},
\bauthor{\bsnm{Matsouaka}, \binits{R.A.}},
\bauthor{\bsnm{Bhatt}, \binits{D.L.}},
\bauthor{\bsnm{Heidenreich}, \binits{P.A.}},
\bauthor{\bsnm{Hernandez}, \binits{A.F.}},
\bauthor{\bsnm{Devore}, \binits{A.D.}},
\bauthor{\bsnm{Yancy}, \binits{C.W.}},
\bauthor{\bsnm{Fonarow}, \binits{G.C.}}:
\batitle{Heart failure with preserved, borderline, and reduced ejection fraction: 5-year outcomes}.
\bjtitle{Journal of the American College of Cardiology}
\bvolume{70}(\bissue{20}),
\bfpage{2476}--\blpage{2486}
(\byear{2017})
\end{barticle}
\endbibitem

\bibitem[\protect\citeauthoryear{Birkeland et~al.}{2020}]{birkeland2020heart}
\begin{barticle}
\bauthor{\bsnm{Birkeland}, \binits{K.I.}},
\bauthor{\bsnm{Bodegard}, \binits{J.}},
\bauthor{\bsnm{Eriksson}, \binits{J.W.}},
\bauthor{\bsnm{Norhammar}, \binits{A.}},
\bauthor{\bsnm{Haller}, \binits{H.}},
\bauthor{\bsnm{Linssen}, \binits{G.C.}},
\bauthor{\bsnm{Banerjee}, \binits{A.}},
\bauthor{\bsnm{Thuresson}, \binits{M.}},
\bauthor{\bsnm{Okami}, \binits{S.}},
\bauthor{\bsnm{Garal-Pantaler}, \binits{E.}}, \betal:
\batitle{Heart failure and chronic kidney disease manifestation and mortality risk associations in type 2 diabetes: a large multinational cohort study}.
\bjtitle{Diabetes, obesity and metabolism}
\bvolume{22}(\bissue{9}),
\bfpage{1607}--\blpage{1618}
(\byear{2020})
\end{barticle}
\endbibitem

\bibitem[\protect\citeauthoryear{Weng et~al.}{2023}]{weng2023effect}
\begin{barticle}
\bauthor{\bsnm{Weng}, \binits{S.-C.}},
\bauthor{\bsnm{Lin}, \binits{C.-F.}},
\bauthor{\bsnm{Hsu}, \binits{C.-Y.}},
\bauthor{\bsnm{Lin}, \binits{S.-Y.}}:
\batitle{Effect of frailty, physical performance, and chronic kidney disease on mortality in older patients with diabetes: a retrospective longitudinal cohort study}.
\bjtitle{Diabetology \& Metabolic Syndrome}
\bvolume{15}(\bissue{1}),
\bfpage{7}
(\byear{2023})
\end{barticle}
\endbibitem

\bibitem[\protect\citeauthoryear{Strain et~al.}{2021}]{strain2021diabetes}
\begin{barticle}
\bauthor{\bsnm{Strain}, \binits{W.D.}},
\bauthor{\bsnm{Down}, \binits{S.}},
\bauthor{\bsnm{Brown}, \binits{P.}},
\bauthor{\bsnm{Puttanna}, \binits{A.}},
\bauthor{\bsnm{Sinclair}, \binits{A.}}:
\batitle{Diabetes and frailty: an expert consensus statement on the management of older adults with type 2 diabetes}.
\bjtitle{Diabetes Therapy}
\bvolume{12}(\bissue{5}),
\bfpage{1227}--\blpage{1247}
(\byear{2021})
\end{barticle}
\endbibitem

\bibitem[\protect\citeauthoryear{Heidenreich et~al.}{2013}]{heidenreich2013forecasting}
\begin{barticle}
\bauthor{\bsnm{Heidenreich}, \binits{P.A.}},
\bauthor{\bsnm{Albert}, \binits{N.M.}},
\bauthor{\bsnm{Allen}, \binits{L.A.}},
\bauthor{\bsnm{Bluemke}, \binits{D.A.}},
\bauthor{\bsnm{Butler}, \binits{J.}},
\bauthor{\bsnm{Fonarow}, \binits{G.C.}},
\bauthor{\bsnm{Ikonomidis}, \binits{J.S.}},
\bauthor{\bsnm{Khavjou}, \binits{O.}},
\bauthor{\bsnm{Konstam}, \binits{M.A.}},
\bauthor{\bsnm{Maddox}, \binits{T.M.}}, \betal:
\batitle{Forecasting the impact of heart failure in the united states: a policy statement from the american heart association}.
\bjtitle{Circulation: Heart Failure}
\bvolume{6}(\bissue{3}),
\bfpage{606}--\blpage{619}
(\byear{2013})
\end{barticle}
\endbibitem

\bibitem[\protect\citeauthoryear{}{2024}]{american2024introduction}
\begin{botherref}
Introduction and methodology: Standards of Care in Diabetes—2024.
American Diabetes Association
(2024)
\end{botherref}
\endbibitem

\bibitem[\protect\citeauthoryear{Aguilar et~al.}{2009}]{aguilar2009relationship}
\begin{barticle}
\bauthor{\bsnm{Aguilar}, \binits{D.}},
\bauthor{\bsnm{Bozkurt}, \binits{B.}},
\bauthor{\bsnm{Ramasubbu}, \binits{K.}},
\bauthor{\bsnm{Deswal}, \binits{A.}}:
\batitle{Relationship of hemoglobin a1c and mortality in heart failure patients with diabetes}.
\bjtitle{Journal of the American College of Cardiology}
\bvolume{54}(\bissue{5}),
\bfpage{422}--\blpage{428}
(\byear{2009})
\end{barticle}
\endbibitem

\bibitem[\protect\citeauthoryear{Evans et~al.}{2010}]{evans2010effect}
\begin{barticle}
\bauthor{\bsnm{Evans}, \binits{J.M.}},
\bauthor{\bsnm{Doney}, \binits{A.S.}},
\bauthor{\bsnm{AlZadjali}, \binits{M.A.}},
\bauthor{\bsnm{Ogston}, \binits{S.A.}},
\bauthor{\bsnm{Petrie}, \binits{J.R.}},
\bauthor{\bsnm{Morris}, \binits{A.D.}},
\bauthor{\bsnm{Struthers}, \binits{A.D.}},
\bauthor{\bsnm{Wong}, \binits{A.K.}},
\bauthor{\bsnm{Lang}, \binits{C.C.}}:
\batitle{Effect of metformin on mortality in patients with heart failure and type 2 diabetes mellitus}.
\bjtitle{The American journal of cardiology}
\bvolume{106}(\bissue{7}),
\bfpage{1006}--\blpage{1010}
(\byear{2010})
\end{barticle}
\endbibitem

\bibitem[\protect\citeauthoryear{Zhou et~al.}{2023}]{zhou2023impact}
\begin{barticle}
\bauthor{\bsnm{Zhou}, \binits{Q.}},
\bauthor{\bsnm{Yang}, \binits{J.}},
\bauthor{\bsnm{Wang}, \binits{W.}},
\bauthor{\bsnm{Shao}, \binits{C.}},
\bauthor{\bsnm{Hua}, \binits{X.}},
\bauthor{\bsnm{Tang}, \binits{Y.-D.}}:
\batitle{The impact of the stress hyperglycemia ratio on mortality and rehospitalization rate in patients with acute decompensated heart failure and diabetes}.
\bjtitle{Cardiovascular Diabetology}
\bvolume{22}(\bissue{1}),
\bfpage{189}
(\byear{2023})
\end{barticle}
\endbibitem

\bibitem[\protect\citeauthoryear{Si et~al.}{2025}]{si2025retrospective}
\begin{botherref}
\oauthor{\bsnm{Si}, \binits{Y.}},
\oauthor{\bsnm{Sun}, \binits{L.}},
\oauthor{\bsnm{Chen}, \binits{S.}},
\oauthor{\bsnm{Fan}, \binits{J.}},
\oauthor{\bsnm{Pishgar}, \binits{E.}},
\oauthor{\bsnm{Alaei}, \binits{K.}},
\oauthor{\bsnm{Placencia}, \binits{G.}},
\oauthor{\bsnm{Pishgar}, \binits{M.}}:
Retrospective machine learning approach for forecasting in-hospital death in icu patients after cardiac arrest.
medRxiv,
2025--05
(2025)
\end{botherref}
\endbibitem

\bibitem[\protect\citeauthoryear{Chen et~al.}{2025}]{chen2025predicting}
\begin{botherref}
\oauthor{\bsnm{Chen}, \binits{S.}},
\oauthor{\bsnm{Fan}, \binits{J.}},
\oauthor{\bsnm{Si}, \binits{Y.}},
\oauthor{\bsnm{Sun}, \binits{L.}},
\oauthor{\bsnm{Alaei}, \binits{K.}},
\oauthor{\bsnm{Pishgar}, \binits{E.}},
\oauthor{\bsnm{Placencia}, \binits{G.}},
\oauthor{\bsnm{Pishgar}, \binits{M.}}:
Predicting icu readmission in acute pancreatitis patients using a machine learning-based model with enhanced clinical interpretability.
medRxiv,
2025--05
(2025)
\end{botherref}
\endbibitem

\bibitem[\protect\citeauthoryear{Gao et~al.}{2024}]{gao2024prediction}
\begin{barticle}
\bauthor{\bsnm{Gao}, \binits{J.}},
\bauthor{\bsnm{Lu}, \binits{Y.}},
\bauthor{\bsnm{Ashrafi}, \binits{N.}},
\bauthor{\bsnm{Domingo}, \binits{I.}},
\bauthor{\bsnm{Alaei}, \binits{K.}},
\bauthor{\bsnm{Pishgar}, \binits{M.}}:
\batitle{Prediction of sepsis mortality in icu patients using machine learning methods}.
\bjtitle{BMC Medical Informatics and Decision Making}
\bvolume{24}(\bissue{1}),
\bfpage{228}
(\byear{2024})
\end{barticle}
\endbibitem

\bibitem[\protect\citeauthoryear{Zhao et~al.}{2021}]{zhao2021development}
\begin{barticle}
\bauthor{\bsnm{Zhao}, \binits{Q.-Y.}},
\bauthor{\bsnm{Wang}, \binits{H.}},
\bauthor{\bsnm{Luo}, \binits{J.-C.}},
\bauthor{\bsnm{Luo}, \binits{M.-H.}},
\bauthor{\bsnm{Liu}, \binits{L.-P.}},
\bauthor{\bsnm{Yu}, \binits{S.-J.}},
\bauthor{\bsnm{Liu}, \binits{K.}},
\bauthor{\bsnm{Zhang}, \binits{Y.-J.}},
\bauthor{\bsnm{Sun}, \binits{P.}},
\bauthor{\bsnm{Tu}, \binits{G.-W.}}, \betal:
\batitle{Development and validation of a machine-learning model for prediction of extubation failure in intensive care units}.
\bjtitle{Frontiers in medicine}
\bvolume{8},
\bfpage{676343}
(\byear{2021})
\end{barticle}
\endbibitem

\bibitem[\protect\citeauthoryear{Johnson et~al.}{2020}]{johnson2020mimic}
\begin{botherref}
\oauthor{\bsnm{Johnson}, \binits{A.}},
\oauthor{\bsnm{Bulgarelli}, \binits{L.}},
\oauthor{\bsnm{Pollard}, \binits{T.}},
\oauthor{\bsnm{Horng}, \binits{S.}},
\oauthor{\bsnm{Celi}, \binits{L.A.}},
\oauthor{\bsnm{Mark}, \binits{R.}}:
Mimic-iv.
PhysioNet. Available online at: https://physionet. org/content/mimiciv/1.0/(accessed August 23, 2021),
49--55
(2020)
\end{botherref}
\endbibitem

\bibitem[\protect\citeauthoryear{Pang et~al.}{2022}]{pang2022establishment}
\begin{barticle}
\bauthor{\bsnm{Pang}, \binits{K.}},
\bauthor{\bsnm{Li}, \binits{L.}},
\bauthor{\bsnm{Ouyang}, \binits{W.}},
\bauthor{\bsnm{Liu}, \binits{X.}},
\bauthor{\bsnm{Tang}, \binits{Y.}}:
\batitle{Establishment of icu mortality risk prediction models with machine learning algorithm using mimic-iv database}.
\bjtitle{Diagnostics}
\bvolume{12}(\bissue{5}),
\bfpage{1068}
(\byear{2022})
\end{barticle}
\endbibitem

\bibitem[\protect\citeauthoryear{Pargent et~al.}{2022}]{pargent2022regularized}
\begin{barticle}
\bauthor{\bsnm{Pargent}, \binits{F.}},
\bauthor{\bsnm{Pfisterer}, \binits{F.}},
\bauthor{\bsnm{Thomas}, \binits{J.}},
\bauthor{\bsnm{Bischl}, \binits{B.}}:
\batitle{Regularized target encoding outperforms traditional methods in supervised machine learning with high cardinality features}.
\bjtitle{Computational Statistics}
\bvolume{37}(\bissue{5}),
\bfpage{2671}--\blpage{2692}
(\byear{2022})
\end{barticle}
\endbibitem

\bibitem[\protect\citeauthoryear{Si et~al.}{2025}]{si2025machine}
\begin{botherref}
\oauthor{\bsnm{Si}, \binits{Y.}},
\oauthor{\bsnm{Fan}, \binits{J.}},
\oauthor{\bsnm{Sun}, \binits{L.}},
\oauthor{\bsnm{Chen}, \binits{S.}},
\oauthor{\bsnm{Pishgar}, \binits{E.}},
\oauthor{\bsnm{Alaei}, \binits{K.}},
\oauthor{\bsnm{Placencia}, \binits{G.}},
\oauthor{\bsnm{Pishgar}, \binits{M.}}:
Machine learning-based prediction of mortality in geriatric traumatic brain injury patients.
arXiv preprint arXiv:2505.15850
(2025)
\end{botherref}
\endbibitem

\bibitem[\protect\citeauthoryear{Jentzer et~al.}{2019}]{JENTZER20191994}
\begin{barticle}
\bauthor{\bsnm{Jentzer}, \binits{J.C.}},
\bauthor{\bsnm{Anavekar}, \binits{N.S.}},
\bauthor{\bsnm{Brenes-Salazar}, \binits{J.A.}},
\bauthor{\bsnm{Wiley}, \binits{B.}},
\bauthor{\bsnm{Murphree}, \binits{D.H.}},
\bauthor{\bsnm{Bennett}, \binits{C.}},
\bauthor{\bsnm{Murphy}, \binits{J.G.}},
\bauthor{\bsnm{Keegan}, \binits{M.T.}},
\bauthor{\bsnm{Barsness}, \binits{G.W.}}:
\batitle{Admission braden skin score independently predicts mortality in cardiac intensive care patients}.
\bjtitle{Mayo Clinic Proceedings}
\bvolume{94}(\bissue{10}),
\bfpage{1994}--\blpage{2003}
(\byear{2019})
\doiurl{10.1016/j.mayocp.2019.04.038}
\end{barticle}
\endbibitem

\bibitem[\protect\citeauthoryear{Azushima et~al.}{2023}]{AZUSHIMA20231135}
\begin{barticle}
\bauthor{\bsnm{Azushima}, \binits{K.}},
\bauthor{\bsnm{Kovalik}, \binits{J.-P.}},
\bauthor{\bsnm{Yamaji}, \binits{T.}},
\bauthor{\bsnm{Ching}, \binits{J.}},
\bauthor{\bsnm{Chng}, \binits{T.W.}},
\bauthor{\bsnm{Guo}, \binits{J.}},
\bauthor{\bsnm{Liu}, \binits{J.-J.}},
\bauthor{\bsnm{Nguyen}, \binits{M.}},
\bauthor{\bsnm{Sakban}, \binits{R.B.}},
\bauthor{\bsnm{George}, \binits{S.E.}},
\bauthor{\bsnm{Tan}, \binits{P.H.}},
\bauthor{\bsnm{Lim}, \binits{S.C.}},
\bauthor{\bsnm{Gurley}, \binits{S.B.}},
\bauthor{\bsnm{Coffman}, \binits{T.M.}}:
\batitle{Abnormal lactate metabolism is linked to albuminuria and kidney injury in diabetic nephropathy}.
\bjtitle{Kidney International}
\bvolume{104}(\bissue{6}),
\bfpage{1135}--\blpage{1149}
(\byear{2023})
\doiurl{10.1016/j.kint.2023.08.006}
\end{barticle}
\endbibitem

\bibitem[\protect\citeauthoryear{Vincent and Zambon}{2006}]{vincent2006ali_mods}
\begin{barticle}
\bauthor{\bsnm{Vincent}, \binits{J.-L.}},
\bauthor{\bsnm{Zambon}, \binits{M.}}:
\batitle{Why do patients who have acute lung injury/acute respiratory distress syndrome die from multiple organ dysfunction syndrome? implications for management}.
\bjtitle{Clinics in Chest Medicine}
\bvolume{27}(\bissue{4}),
\bfpage{725}--\blpage{731}
(\byear{2006})
\doiurl{10.1016/j.ccm.2006.06.010}
\end{barticle}
\endbibitem

\bibitem[\protect\citeauthoryear{Shu et~al.}{2023}]{shu2023development}
\begin{barticle}
\bauthor{\bsnm{Shu}, \binits{T.}},
\bauthor{\bsnm{Huang}, \binits{J.}},
\bauthor{\bsnm{Deng}, \binits{J.}},
\bauthor{\bsnm{Chen}, \binits{H.}},
\bauthor{\bsnm{Zhang}, \binits{Y.}},
\bauthor{\bsnm{Duan}, \binits{M.}},
\bauthor{\bsnm{Wang}, \binits{Y.}},
\bauthor{\bsnm{Hu}, \binits{X.}},
\bauthor{\bsnm{Liu}, \binits{X.}}:
\batitle{Development and assessment of scoring model for icu stay and mortality prediction after emergency admissions in ischemic heart disease: a retrospective study of mimic-iv databases}.
\bjtitle{Internal and Emergency Medicine}
\bvolume{18}(\bissue{2}),
\bfpage{487}--\blpage{497}
(\byear{2023})
\end{barticle}
\endbibitem

\bibitem[\protect\citeauthoryear{Dou et~al.}{2023}]{dou2023association}
\begin{barticle}
\bauthor{\bsnm{Dou}, \binits{J.}},
\bauthor{\bsnm{Guo}, \binits{C.}},
\bauthor{\bsnm{Wang}, \binits{Y.}},
\bauthor{\bsnm{Peng}, \binits{Z.}},
\bauthor{\bsnm{Wu}, \binits{R.}},
\bauthor{\bsnm{Li}, \binits{Q.}},
\bauthor{\bsnm{Zhao}, \binits{H.}},
\bauthor{\bsnm{Song}, \binits{S.}},
\bauthor{\bsnm{Sun}, \binits{X.}},
\bauthor{\bsnm{Wei}, \binits{J.}}:
\batitle{Association between triglyceride glucose-body mass and one-year all-cause mortality of patients with heart failure: a retrospective study utilizing the mimic-iv database}.
\bjtitle{Cardiovascular diabetology}
\bvolume{22}(\bissue{1}),
\bfpage{309}
(\byear{2023})
\end{barticle}
\endbibitem

\bibitem[\protect\citeauthoryear{Zhang et~al.}{2022}]{zhang2022prediction}
\begin{barticle}
\bauthor{\bsnm{Zhang}, \binits{L.}},
\bauthor{\bsnm{Huang}, \binits{T.}},
\bauthor{\bsnm{Xu}, \binits{F.}},
\bauthor{\bsnm{Li}, \binits{S.}},
\bauthor{\bsnm{Zheng}, \binits{S.}},
\bauthor{\bsnm{Lyu}, \binits{J.}},
\bauthor{\bsnm{Yin}, \binits{H.}}:
\batitle{Prediction of prognosis in elderly patients with sepsis based on machine learning (random survival forest)}.
\bjtitle{BMC emergency medicine}
\bvolume{22}(\bissue{1}),
\bfpage{26}
(\byear{2022})
\end{barticle}
\endbibitem

\bibitem[\protect\citeauthoryear{Yuan et~al.}{2024}]{yuan2024xgboost}
\begin{botherref}
\oauthor{\bsnm{Yuan}, \binits{W.}},
\oauthor{\bsnm{Xiao}, \binits{M.}},
\oauthor{\bsnm{Wang}, \binits{R.}},
\oauthor{\bsnm{Liu}, \binits{G.}},
\oauthor{\bsnm{Wu}, \binits{J.}},
\oauthor{\bsnm{Wang}, \binits{X.}}:
Xgboost in the prediction of 28-day mortality in critical elderly patients with hip fracture: A mimic-iv cohort study.
Alternative Therapies in Health \& Medicine
\textbf{30}(9)
(2024)
\end{botherref}
\endbibitem

\bibitem[\protect\citeauthoryear{Xu et~al.}{2022}]{xu2022predicting}
\begin{barticle}
\bauthor{\bsnm{Xu}, \binits{Y.}},
\bauthor{\bsnm{Han}, \binits{D.}},
\bauthor{\bsnm{Huang}, \binits{T.}},
\bauthor{\bsnm{Zhang}, \binits{X.}},
\bauthor{\bsnm{Lu}, \binits{H.}},
\bauthor{\bsnm{Shen}, \binits{S.}},
\bauthor{\bsnm{Lyu}, \binits{J.}},
\bauthor{\bsnm{Wang}, \binits{H.}}:
\batitle{Predicting icu mortality in rheumatic heart disease: comparison of xgboost and logistic regression}.
\bjtitle{Frontiers in Cardiovascular Medicine}
\bvolume{9},
\bfpage{847206}
(\byear{2022})
\end{barticle}
\endbibitem

\bibitem[\protect\citeauthoryear{Ren et~al.}{2024}]{ren2024prediction}
\begin{barticle}
\bauthor{\bsnm{Ren}, \binits{W.}},
\bauthor{\bsnm{Zou}, \binits{K.}},
\bauthor{\bsnm{Huang}, \binits{S.}},
\bauthor{\bsnm{Xu}, \binits{H.}},
\bauthor{\bsnm{Zhang}, \binits{W.}},
\bauthor{\bsnm{Shi}, \binits{X.}},
\bauthor{\bsnm{Shi}, \binits{L.}},
\bauthor{\bsnm{Zhong}, \binits{X.}},
\bauthor{\bsnm{Peng}, \binits{Y.}},
\bauthor{\bsnm{Tang}, \binits{X.}}, \betal:
\batitle{Prediction of in-hospital mortality of intensive care unit patients with acute pancreatitis based on an explainable machine learning algorithm}.
\bjtitle{Journal of Clinical Gastroenterology}
\bvolume{58}(\bissue{6}),
\bfpage{619}--\blpage{626}
(\byear{2024})
\end{barticle}
\endbibitem

\bibitem[\protect\citeauthoryear{Ding et~al.}{2021}]{ding2021artificial}
\begin{barticle}
\bauthor{\bsnm{Ding}, \binits{N.}},
\bauthor{\bsnm{Guo}, \binits{C.}},
\bauthor{\bsnm{Li}, \binits{C.}},
\bauthor{\bsnm{Zhou}, \binits{Y.}},
\bauthor{\bsnm{Chai}, \binits{X.}}:
\batitle{An artificial neural networks model for early predicting in-hospital mortality in acute pancreatitis in mimic-iii}.
\bjtitle{BioMed research international}
\bvolume{2021}(\bissue{1}),
\bfpage{6638919}
(\byear{2021})
\end{barticle}
\endbibitem

\bibitem[\protect\citeauthoryear{Vrugt et~al.}{2009}]{vrugt2009accelerating}
\begin{barticle}
\bauthor{\bsnm{Vrugt}, \binits{J.A.}},
\bauthor{\bsnm{Ter~Braak}, \binits{C.J.}},
\bauthor{\bsnm{Diks}, \binits{C.G.}},
\bauthor{\bsnm{Robinson}, \binits{B.A.}},
\bauthor{\bsnm{Hyman}, \binits{J.M.}},
\bauthor{\bsnm{Higdon}, \binits{D.}}:
\batitle{Accelerating markov chain monte carlo simulation by differential evolution with self-adaptive randomized subspace sampling}.
\bjtitle{International Journal of Nonlinear Sciences and Numerical Simulation}
\bvolume{10}(\bissue{3}),
\bfpage{273}--\blpage{290}
(\byear{2009})
\end{barticle}
\endbibitem

\end{thebibliography}

\end{document}